\documentclass[%
superscriptaddress,
preprintnumbers,
nofootinbib,
 amsmath,amssymb,
 aps,
 prd,
 twocolumn,
]{revtex4-2}

\usepackage{graphicx}
\usepackage{dcolumn}
\usepackage{bm}
\usepackage{xcolor}
\usepackage{enumitem}

\usepackage{hyperref}
\usepackage[nameinlink]{cleveref}
\usepackage{import}
\usepackage{algorithm}
\usepackage[noend]{algpseudocode}
\usepackage{aas_macros}

\newcommand{\beq}{\begin{equation}}
\newcommand{\eeq}{\end{equation}}

\newcommand{\COSMOLIKE}{\texttt {CosmoLike}}

\newcommand{\Halofit}{{\texttt{Halofit}}}
\newcommand{\balrog}{{\texttt{Balrog}}}

\newcommand{\metacal}{{\texttt{Metacalibration}~}}
\newcommand{\ttt}{{$3\times2$-point}}
\newcommand{\twottwo}{{$2\times2$-point}}

\usepackage{xspace}

\newcommand{\sdir}{\texttt{3sDir}\xspace}

\newcommand{\Nsims}{{{18}}}
\renewcommand{\thefootnote}{\arabic{footnote}}

\newcommand{\LCDM}{$ \Lambda $CDM~}

\newcommand{\buzzard}{\texttt{Buzzard}}
\newcommand{\addgals}{\texttt{Addgals}}
\newcommand{\calclens}{\texttt{Calclens}}
\newcommand{\buzzardtwo}{\texttt{Buzzard v2.0}}
\newcommand{\redmagic}{\texttt{redMaGiC}}
\newcommand{\maglim}{\texttt{maglim}}
\newcommand{\kcorrect}{\texttt{kcorrect}}
\newcommand{\cosmosis}{\texttt{cosmosis}}

\newcommand{\mpc}{\ensuremath{h^{-1}\mathrm{Mpc}}}

\newcommand{\gpcc}{\ensuremath{h^{-3}\mathrm{Gpc}^3}}

\newcommand{\nbody}{$N$-body}
\newcommand{\wtheta}{\ensuremath{w(\theta)}}
\newcommand{\gammat}{\ensuremath{\gamma_t(\theta)}}
\newcommand{\cs}{\ensuremath{\xi_{\pm}(\theta)}}
\newcommand{\xipm}{\ensuremath{\xi_{\pm}(\theta)}}
\newcommand{\seight}{\ensuremath{S_8}}
\newcommand{\om}{\ensuremath{\Omega_m}}

\newcommand{\zbar}{\bar{z}}

\newcommand{\photoz}{photo-$z$}

\begin{document}
\preprint{DES-2021-0558}
\preprint{FERMILAB-PUB-21-223-AE}

\title{Dark Energy Survey Year 3 results: cosmology from combined galaxy clustering and lensing – validation on cosmological simulations}

\author{J.~DeRose}\email{Corresponding author: jderose@lbl.gov}
\affiliation{Lawrence Berkeley National Laboratory, 1 Cyclotron Road, Berkeley, CA 94720, USA}
\author{R.~H.~Wechsler}
\affiliation{Department of Physics, Stanford University, 382 Via Pueblo Mall, Stanford, CA 94305, USA}
\affiliation{Kavli Institute for Particle Astrophysics \& Cosmology, P. O. Box 2450, Stanford University, Stanford, CA 94305, USA}
\affiliation{SLAC National Accelerator Laboratory, Menlo Park, CA 94025, USA}
\author{M.~R.~Becker}
\affiliation{Argonne National Laboratory, 9700 South Cass Avenue, Lemont, IL 60439, USA}
\author{E.~S.~Rykoff}
\affiliation{Kavli Institute for Particle Astrophysics \& Cosmology, P. O. Box 2450, Stanford University, Stanford, CA 94305, USA}
\affiliation{SLAC National Accelerator Laboratory, Menlo Park, CA 94025, USA}
\author{S.~Pandey}
\affiliation{Department of Physics and Astronomy, University of Pennsylvania, Philadelphia, PA 19104, USA}
\author{N.~MacCrann}
\affiliation{Department of Applied Mathematics and Theoretical Physics, University of Cambridge, Cambridge CB3 0WA, UK}
\author{A.~Amon}
\affiliation{Kavli Institute for Particle Astrophysics \& Cosmology, P. O. Box 2450, Stanford University, Stanford, CA 94305, USA}
\author{J.~Myles}
\affiliation{Department of Physics, Stanford University, 382 Via Pueblo Mall, Stanford, CA 94305, USA}
\affiliation{Kavli Institute for Particle Astrophysics \& Cosmology, P. O. Box 2450, Stanford University, Stanford, CA 94305, USA}
\affiliation{SLAC National Accelerator Laboratory, Menlo Park, CA 94025, USA}
\author{E.~Krause}
\affiliation{Department of Astronomy/Steward Observatory, University of Arizona, 933 North Cherry Avenue, Tucson, AZ 85721-0065, USA}
\author{D.~Gruen}
\affiliation{Department of Physics, Stanford University, 382 Via Pueblo Mall, Stanford, CA 94305, USA}
\affiliation{Kavli Institute for Particle Astrophysics \& Cosmology, P. O. Box 2450, Stanford University, Stanford, CA 94305, USA}
\affiliation{SLAC National Accelerator Laboratory, Menlo Park, CA 94025, USA}
% ...

\collaboration{DES Collaboration}

\date{\today}

\begin{abstract}
We present a validation of the Dark Energy Survey Year 3 (DES Y3) $3\times2$-point analysis choices by testing them on \buzzardtwo, a new suite of cosmological simulations that is tailored for the testing and validation of combined galaxy clustering and weak lensing analyses. We show that the \buzzardtwo\ simulations accurately reproduce many important aspects of the DES Y3 data, including photometric redshift and magnitude distributions, and the relevant set of two-point clustering and weak lensing statistics. We then show that our model for the \ttt\ data vector is accurate enough to recover the true cosmology in simulated surveys assuming the true redshift distributions for our source and lens samples, demonstrating robustness to uncertainties in the modeling of the non-linear matter power spectrum, non-linear galaxy bias and higher-order lensing corrections. Additionally, we demonstrate for the first time that our photometric redshift calibration methodology, including information from photometry, spectroscopy, clustering cross-correlations and galaxy--galaxy lensing ratios, is accurate enough to recover the true cosmology in simulated surveys in the presence of realistic photometric redshift uncertainties.
\end{abstract}

\maketitle

\section{Introduction}
Having effectively exhausted the information encoded in the linear modes probed by the Cosmic Microwave Background (CMB), cosmologists have lately turned to wide-field galaxy surveys as their preferred tool for studying dark matter and dark energy. These galaxy surveys, which probe the distribution of matter in the low-redshift universe, encode a great deal of cosmological information, but their complex observational and modeling systematics pose significant challenges to unbiased inference. 

Recently, combinations of galaxy clustering and weak lensing have been realized as powerful mechanisms for extracting cosmological information from photometric galaxy surveys such as the Dark Energy Survey \footnote{https://www.darkenergysurvey.org/} (DES), Kilo Degree Survey \footnote{http://kids.strw.leidenuniv.nl/} (KiDS), and Hyper Suprime Cam \footnote{https://hsc.mtk.nao.ac.jp/ssp/} (HSC). In particular, the combination of shear--shear (cosmic shear, \xipm), galaxy position--galaxy position (galaxy clustering, \wtheta) and tangential shear--galaxy position (galaxy--galaxy lensing, \gammat) two-point functions into a \ttt\ analysis has proven powerful, in part because of its ability to break degeneracies between nuisance parameters and cosmological parameters \citep{y1kp, Heymans2020}. The resultant increase in constraining power comes at a cost though: great care must be taken to ensure that all of the components that feed into the \ttt\ analysis are robustly determined, lest the inferred cosmological constraints be biased.

While some analysis validation can be undertaken analytically \citep{methodpaper}, or using the data itself \citep{Huff2015}, validation against realistic cosmological simulations is an essential component of modern galaxy survey cosmology analyses \citep{MacCrann2019, Sugiyama2020, Rossi2020, Pandey2020}. In this work, we validate three main components of the \ttt\ analysis being performed on the first three years of DES data (DES Y3): the two-point function measurement pipeline, photometric redshift calibration methodology and the likelihood and modeling framework used to obtain cosmological constraints from two-point functions and redshift distributions.

The first challenge that must be overcome in \ttt\ analyses is the accurate measurement of the cosmic shear, galaxy--galaxy lensing and galaxy angular clustering statistics that make up the \ttt\ data vector. The measurement of galaxy ellipticities is especially important for weak lensing analyses, and dedicated image simulations that test for and constrain biases that appear in this process are an essential ingredient for modern weak lensing analyses \cite{Mandelbaum2017,Kannawadi2018,Samuroff2018,MacCrann2020,y3-balrog,y3-shapecatalog}. In this work, we focus on ensuring that our two-point measurement pipelines deliver unbiased correlation functions in the absence of these shear systematics. Mitigating angular systematics imprinted on galaxy clustering and galaxy--galaxy lensing statistics by survey and foreground inhomogeneities is also important, and while we do imprint DES depth variations on our simulations that lead to systematics of this kind, the DES Y3 methodology for removing such systematics is thoroughly investigated independently in \citep{wthetapaper, y3-galaxyclustering, y3-gglensing}.

The next key component required for accurate cosmological inference is a characterization of the source and lens galaxy sample redshift distributions. The robust calibration of source redshift distributions is a challenge shared by all weak lensing surveys \cite{Tanaka2017,Hildebrant2020,y3-sompz}. A critical component of such a calibration is correctly accounting for incompleteness of the spectroscopic galaxy catalogs that form the foundation of all photometric redshift estimation. Previous versions of the simulations presented in this work have been used to study these effects \citep{Hartley2020}, but here we assume that we have complete, albeit realistically sized, spectroscopic redshift catalogs, upon which we build our photometric redshift estimations. We proceed to test the three separate components of the DES Y3 photometric redshift calibration: photometric redshift estimation using self-organizing maps (SOMPZ and 3sDir), clustering redshifts (WZ), and the ratios of galaxy--galaxy lensing signals (shear ratios). These three components are thoroughly validated in \citet*{y3-sompz,y3-sourcewz,y3-shearratio,Sanchez2020, y3-cosmicshear1}. In this work we present further tests showing that the combination of these three methods is self-consistent and unbiased in our simulations, and for the first time show that the source redshift distributions calibrated using these methods deliver unbiased cosmological constraints in simulations.

The challenge of lens galaxy photometric redshift estimation is also formidable. The lens sample that we use in this work, \redmagic\,, is a sample of luminous red galaxies selected to be constant in comoving number density as a function of redshift. As they are a bright subset of galaxies, photometric redshift estimation is relatively straight-forward compared to that for source galaxy, but nonetheless must be thoroughly validated. In the DES Y3 data, clustering redshifts are again used for this task, and this methodology is tested using the simulations presented here in \citet{y3-lenswz}. The \redmagic\ redshift estimation is further validated in this work, where we show that use of \redmagic\ photometric redshift distributions does not bias our cosmological inference. We note that after un-blinding the DES Y3 \redmagic\ analysis, the fiducial DES Y3 analysis was shifted to use a different lens galaxy sample, called the \maglim\ sample \citep{y3-2x2maglimforecast, y3-2x2ptaltlensresults}. Validation of that analysis, including the use of one of the simulations, described in this work is presented in \citet{y3-2x2ptaltlensresults}.

With robust two-point function and redshift distribution measurements in hand, one remaining challenge is to accurately predict the dependence of the \ttt\ data vector on the cosmological parameters of interest. Weak lensing is particularly sensitive to non-linearities induced by gravitational collapse during the process of large-scale structure formation, and so validation of the assumed matter power spectrum model is a particularly important aspect of \ttt\ analyses. The $N$-body simulations used in this work were run with different settings and resolutions to those used in the \Halofit\ model for the matter power spectrum that the DES Y3 \ttt\ analysis employs, and thus provide an independent validation of this modeling alongside that presented in \citet{y3-generalmethods}. The ray-tracing algorithm that we employ \citep{Becker2013} allows us to incorporate higher-order lensing effects such as magnification, angular deflection, and reduced shear in the two-point functions that we measure in our simulations, validating the approximations of these effects that are made in our modeling framework that are also validated in \citet{y3-generalmethods,y3-2x2ptmagnification,y3-gglensing}.

Robust marginalization over astrophysical nuisance parameters such as galaxy bias, redshift-space distortions (RSD), galaxy intrinsic alignments, and baryonic effects on the matter distribution is another important challenge when interpreting the measurements made in \ttt\ analyses. As the models used for these effects in the DES Y3 \ttt\ analysis are largely perturbative in nature, it is essential to validate them against fully non-linear solutions for these physical processes as implemented in cosmological simulations. In this work, we focus on testing our models for galaxy bias and RSD, while \citet{y3-generalmethods} and \citet*{y3-cosmicshear2} test our models for intrinsic alignments and baryonic effects. These tests complement those performed in \citet{Pandey2020a}, further validating the bias modeling choices that are used in the analysis of the DES Y3 data in \citet{y3-2x2ptbiasmodelling} and the accompanying \twottwo\ analyses. In particular, we show here that the scale cuts used for our analyses are robust to non-linearities beyond those assumed in our model, by showing that the cosmological constraints obtained from the measurements on our simulations are unbiased at high significance. 

Our presentation will be organized as follows. In \cref{sec:buzzsum} we describe the \buzzardtwo\ simulations, which represent a significant upgrade over the simulations used in DES Year 1 analyses. In \cref{sec:datavec} we describe how \ttt\ measurements are made and how redshift distributions are estimated in our simulated analyses. In \cref{sec:model} we describe the model used to obtain cosmological constraints from the DES Y3 \ttt\ measurements that we wish to test on these simulations. In \cref{sec:validation} we present the results of our simulated analyses, including varying levels of realism. Finally, in \cref{sec:conclusion} we summarize our work and conclude by discussing future directions for improvement.

\section{\texttt{Buzzard v2.0} Simulations}
\label{sec:buzzsum}
In this work, we make use of a suite of \Nsims\ \nbody\ simulations that are designed to reproduce the lens and source samples used in the DES Y3 \ttt\ analysis. These are a new version of the \texttt{Buzzard} simulations \citep{DeRose2019}, implementing a number of improvements over those used in analyses of the first year of DES data (DES Y1). In this section we briefly summarize the pertinent details from \citet{DeRose2019}, and outline the main improvements, relegating a more detailed description of these changes to App. \ref{app:colordepclustering}-\ref{app:errormodel}. The main improvements of the simulations presented in this work over those used in DES Y1 analyses are
\begin{enumerate}
    \item Improved color-dependent clustering, based on a conditional abundance matching model.
    \item Explicit matching of red-sequence color distributions to the DES data.
    \item More realistic photometric errors enabled by the \balrog\ \citep{y3-balrog} image simulation framework.
\end{enumerate}

Each \buzzard\ simulation is constructed from three independent \nbody\ simulations, with sizes of $1.05^3$, $2.6^3$, and $4.0^3$ \gpcc\ and particle loads of $1400^3$, $2048^3$, and $2048^3$ respectively. All simulations were run using the \texttt{L-Gadget2} code \citep{Springel2005}, and initialized with independent seeds at $z=50$ using 2nd-order Lagrangian perturbation theory as implemented in \texttt{2LPTIC} \citep{Crocce2006} and linear power spectra computed with \textrm{CAMB} \citep{Lewis2000}, assuming a flat \LCDM\ cosmology with $\Omega_{m}=0.286$, $\Omega_{b}=0.046$, $h=0.7$, $n_s=0.96$, and $\sigma_8=0.82$. A single set of these simulations is sufficient to generate a unique lightcone with an area of $10,413$ square degrees out to $z=2.35$. For further details regarding simulation specifications and validation of relevant observables see Sections 3.1 and 3.2 in \citet{DeRose2019}. 

Galaxies are included in these lightcones using the \addgals\ algorithm \citep{Wechsler2021}, which imbues each galaxy with a position, velocity, size, ellipticity, and spectral energy distribution. Functionally, the algorithm for producing our source and lens galaxy samples proceeds as follows:

\begin{enumerate}
    \item Assign galaxy positions, velocities and absolute magnitudes.
    \item Assign galaxy SEDs.
    \item Assign intrinsic galaxy sizes and ellipticities.
    \item Perform ray-tracing and lensing of the galaxy catalog.
    \item Generate survey-specific photometry and ellipticities.
    \item Select source and lens galaxy samples.
\end{enumerate}

Step 1 is unchanged from previous versions of the \buzzard\ simulations, and is fully explained in \citet{Wechsler2021}. We generate a list of galaxy absolute magnitudes by integrating a luminosity function that has been fit to a variety of spectroscopic data sets. Additionally, we have optimized the luminosity function in order to match observed galaxy counts as a function of magnitude in the DES Y3 data, using an algorithm analogous to that used in DES Y1, described in Appendix E.1 of \citet{DeRose2019}. Galaxies from this list are assigned phase-space coordinates using a model for $P(\delta|M_{r},z)$ that is tuned to a subhalo abundance matching (SHAM) model, where $P(\delta|M_{r},z)$ is the probability that a galaxy in the simulation with an absolute magnitude $M_r$ is found in an overdensity $\delta$. In practice $\delta$ is defined as the inverse of the radius enclosing a $3\times10^{13}\, h^{-1}M_{\odot}$. In halo masses around and below this mass, \addgals\ tends to be slightly less accurate as described in \citet{Wechsler2021}. 

One of the main galaxy samples used for DES cosmological constraints is the \redmagic\ sample, which is designed to have robust photometric redshifts made possible by preferentially selecting bright red galaxies whose redshift distributions have been shown to be accurately and precisely characterized \citep{Rozo2015,Cawthon2018,y3-lenswz}. In order to better model this sample, we have made significant improvements to the color-dependent clustering model with respect to previous versions of these simulations. In particular we adopt a conditional abundance matching algorithm that assigns SEDs from the SDSS Main Galaxy Sample (SDSS MGS) with redder rest frame $g-r$ colors to galaxies that are closer to dark matter halos with masses greater than $M_{\rm 200b} \sim 10^{13}\, h^{-1} M_{\odot}$. A more detailed description of this algorithm can be found in App. \ref{app:colordepclustering} as well as in \citet{Wechsler2021} and \citet{DeRose2021}. 

\begin{figure}
	\includegraphics[width=\columnwidth]{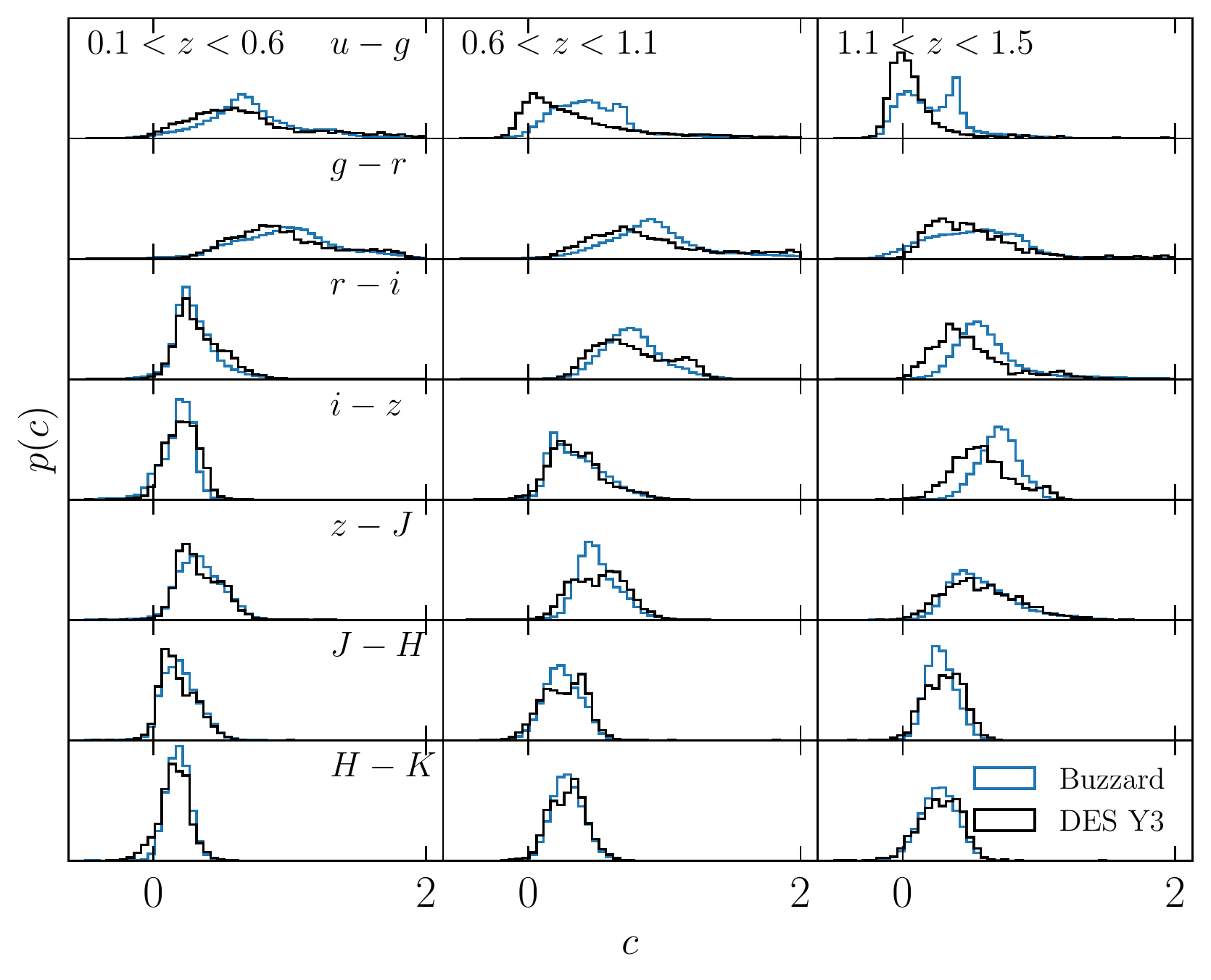}
    \caption{Comparison of $ugrizJHK$ color distributions as a function of redshift between Buzzard (blue) and the DES Y3 redshift sample (black). Different rows depict color distributions for different band combinations (listed in left column), while different columns show different redshift bins. Agreement is good, except for in the $u$-band and at redshifts $z>1$, where the SED templates used in Buzzard are poorly constrained by data \citep{Blanton2005}.}
    \label{fig:colors}
\end{figure}

Following this procedure, we compute broad-band magnitudes by integrating each SED over DES $ugrizY$ and VISTA $JHK$ band-passes. This procedure leads to color distributions that well approximate those observed in the DES Y3 data. In particular, color distributions for \buzzard\ and the DES Y3 deep field data are compared in \cref{fig:colors}, where we have binned the deep field data in redshift using the most precise redshift estimate available for each galaxy. Agreement is excellent except for in $u-g$, where there are known deficiencies in the SED templates used in our simulations \citep{DeRose2019}. 

\begin{figure}
	\includegraphics[width=\columnwidth]{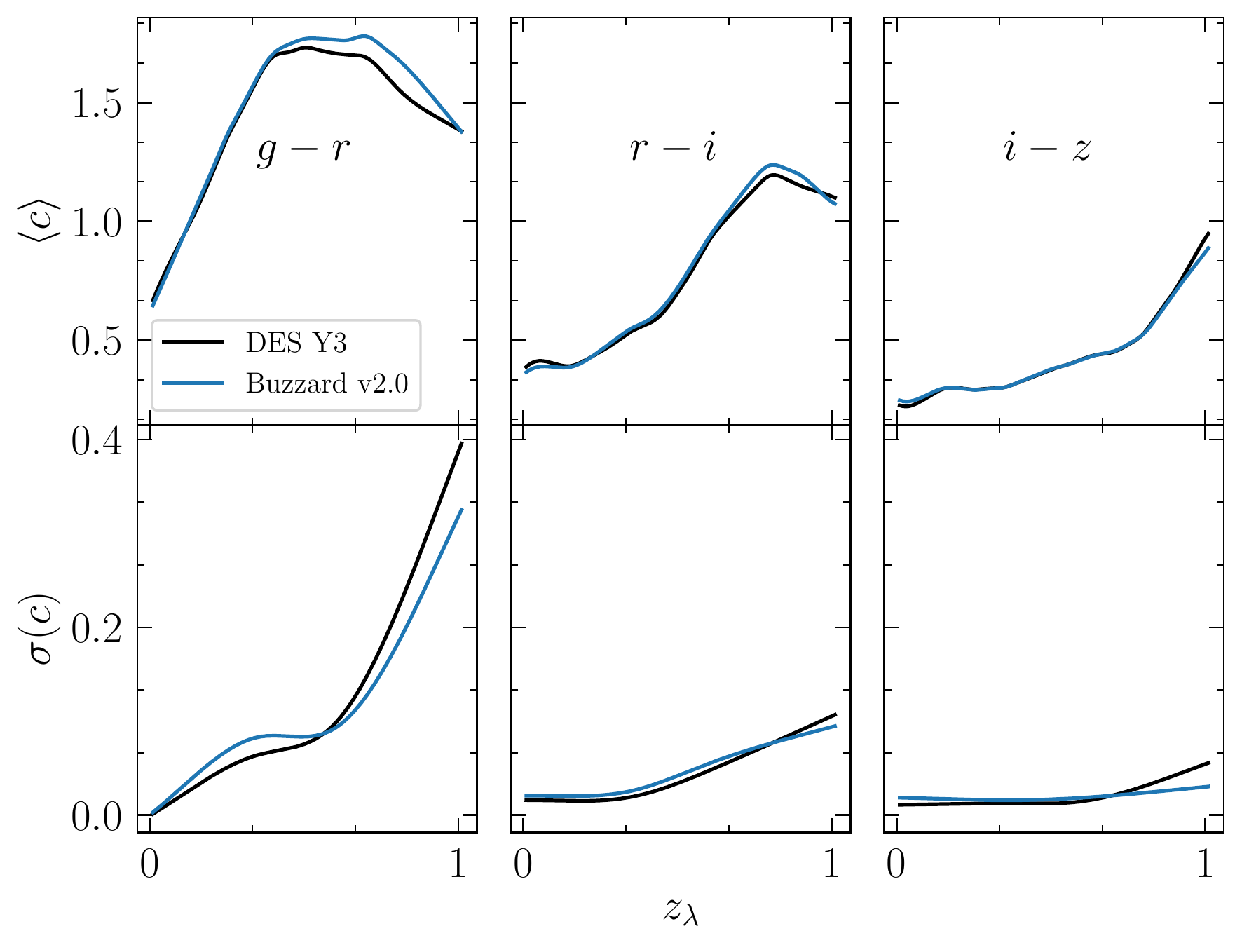}
    \caption{Comparison of red-sequence colors between Buzzard and DES Y3 as measured by \texttt{redMaPPer}. ({\it Top}) Mean red-sequence color as a function of redshift for the DES Y3 data (black) compared to the \buzzard\ simulations (blue). The largest differences occur at high redshift in $g-r$, where the mean color is poorly constrained in the data. ({\it Bottom}) Scatter in red-sequence colors as a function of redshift for the DES Y3 data and \buzzard.}
    \label{fig:red_sequence}
\end{figure}

In \citet{DeRose2019} we showed that the galaxy red sequence, e.g. $P(r-i | z)$ for red galaxies, in the DES Y1 \buzzard\ simulations was significantly narrower than that found in the DES Y1 data. This led to over-optimistic photometric redshift uncertainties for red-sequence galaxies such as the \redmagic\ sample, as the uncertainty in these photometric redshift estimates is directly proportional to the width of the red-sequence. Additionally, at high redshifts, galaxies in the DES Y1 \buzzard\ simulations were significantly redder than those found in the data, leading to a deficit of bright, red galaxies. In order to remedy these issues, we explicitly force the mean and width of the red-sequence to match that found in the data. The agreement between the \buzzardtwo\ simulations and the DES Y3 data in this respect is shown in \cref{fig:red_sequence}. The algorithm for performing this matching is described in App. \ref{app:redsequence}.

Once SEDs have been assigned and magnitudes generated in each DES band-pass, we assign galaxy half-light radii and intrinsic ellipticities as a function of observed $i$-band magnitude as described in App. E3 in \citet{DeRose2019}. After this step we have a catalog of galaxies with true positions, velocities, SEDs, DES magnitudes sizes, and ellipticities.

Before applying any survey-specific masking or photometric errors to our simulations, we compute weak-lensing quantities by ray-tracing through our simulations with the \calclens\ code \citep{Becker2013} using the same configuration as described in \citet{DeRose2019}. We then compute deflection, rotation, shear, and magnification for each galaxy from the lensing distortion tensor at the position of that galaxy. These quantities are then used to deflect the angular positions, rotate and shear the intrinsic ellipticities, and magnify the sizes and magnitudes of each galaxy.

After lensing has been performed, the simulations are rotated into the DES Y3 footprint, and masked using the DES Y3 \redmagic\ mask. We are able to cut two DES Y3 footprints per $10,000$ square degree lightcone, each with an area of 4143.17 square degrees. 

In addition to these improvements, we have implemented a more realistic model for photometric errors in order to provide a better testing ground for the photometric redshift methodologies employed in the DES Y3 \ttt\ analysis. Along with the Gaussian photometric errors that we produce using the model described in Appendix E.4 of \citet{DeRose2019}, we produce independent realizations of photometric errors by making use of the photometric error distributions measured by the \balrog\ image simulation framework \citep{Huff2015,y3-balrog}. \balrog\ injects low-noise images of galaxies observed in the DES deep fields \citep{y3-deepfields} into wide-field images and remeasures their photometry. This allows for estimates of detection efficiency and photometric errors as a function of the nearly noiseless deep-field galaxy properties. These relationships are precisely what are required in order to apply photometric errors to our simulations. For each simulated galaxy, we find a \balrog\ injected galaxy that is closest in $riz$ bands and apply the magnitude offsets between the true \balrog\ injected photometry and the measured wide-field detection. For a detailed description of this algorithm, see App. \ref{app:errormodel}. 
At this point we have galaxy catalogs with lensed positions, velocities, SEDs, sizes, ellipticities, and $ugrizYJHK$ magnitudes with realistic noise from which we can select DES-like galaxy samples.

\section{Measurement of the \ttt\ Data Vector}
\label{sec:datavec}
The \ttt\ data vector consists of the combination of galaxy ellipticity and galaxy density auto correlations, the cross correlation between galaxy density and galaxy ellipticities, and the associated source and lens galaxy redshift distributions. In our simulations we have kept as close as possible to the measurement pipelines used in the DES Y3 data, with the main exception being that our simulated source catalog does not have inverse variance weights or \metacal shear responses \citep{Huff2015,Sheldon2017}. We have also opted to measure all shear correlation functions without shape noise in order to maximize the constraining power afforded us by this suite of simulations. \Cref{sec:sample_selection} describes how we select source and lens galaxy catalogs from the \buzzard\ simulations and \Cref{sec:twopoint_est} describes the estimators used to construct the two-point functions that make up the \ttt\ data vector. 

\begin{figure*}
	\includegraphics[width=\linewidth]{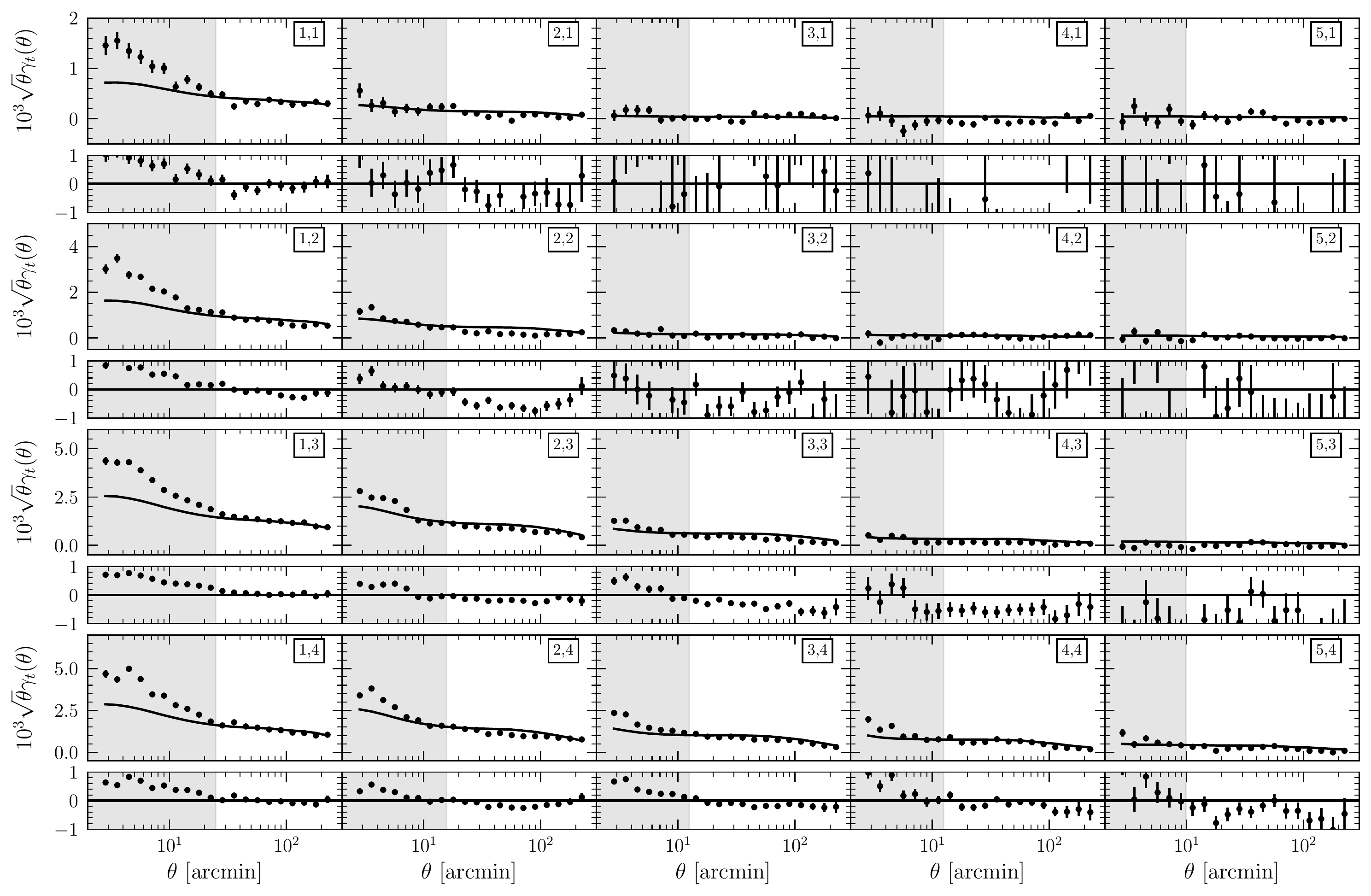}
	\includegraphics[width=\linewidth]{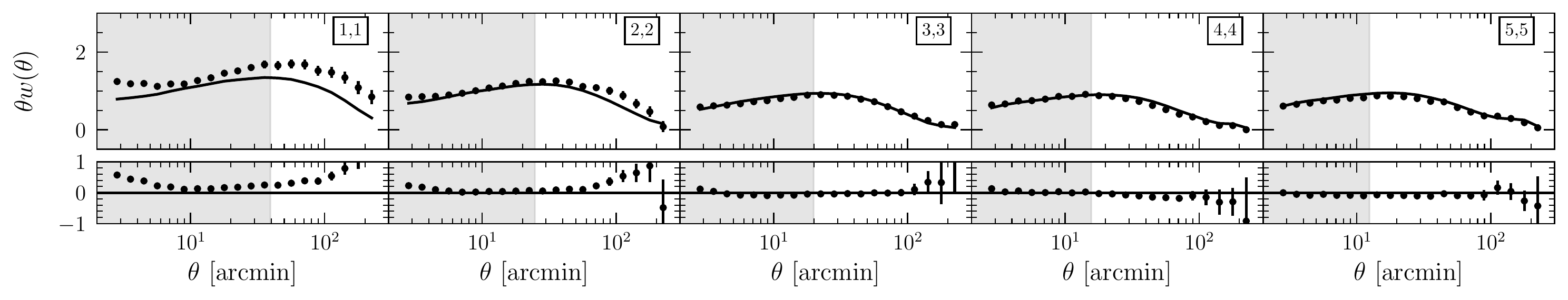}
    \caption{Comparison of \gammat\ and \wtheta\ between \buzzard\ (lines) and the measurements from the Y3 data using the \redmagic\ lens sample (points), with error bars given by the fiducial Y3 covariance matrix. Shaded regions indicate the angular scale cuts applied in the fiducial \ttt\ analysis. Rows alternate between showing the measured signals and the fractional difference between data and simulations. Numbers in each panel label the bin combinations shown, with the numbers in the \gammat\ panels representing the lens-source bin pair, and numbers in \wtheta\ panels denoting the lens bins alone. The intent of this comparison is to gauge how well the galaxy clustering and lensing properties in \buzzard\ match those observed in the data, with the caveat that disagreement may arise due to differences in cosmology, and source and lens galaxy redshift distributions. Agreement in \wtheta\ is generally better than for \gammat\, especially for the 2nd--5th lens bins, which may be a result of slightly different source redshift distributions between the simulations and data, especially where there is significant overlap between source and lens redshift distributions. We also observe a marked deficit of power on small scales in \gammat\ for the first lens bin at smaller scales than those used in the analysis, part of which is likely due to resolution effects in \buzzard. The excellent match between \gammat\ on larger scales in the first lens bin is interesting in light of the large discrepancy in the amplitude of \wtheta\ between the DES Y3 data and \buzzard\ for this lens bin. \citet{y3-3x2ptkp} demonstrates that under the assumption of LCDM the \redmagic\ \wtheta\ and \gammat\ imply bias values that differ by $12\%$, so this may play some role in the discrepancies seen between the \buzzard\ and DES Y3 \wtheta\ measurements in the first lens bin.}
    \label{fig: 2x2_data_sims}
\end{figure*}

\begin{figure*}
	\includegraphics[width=\linewidth]{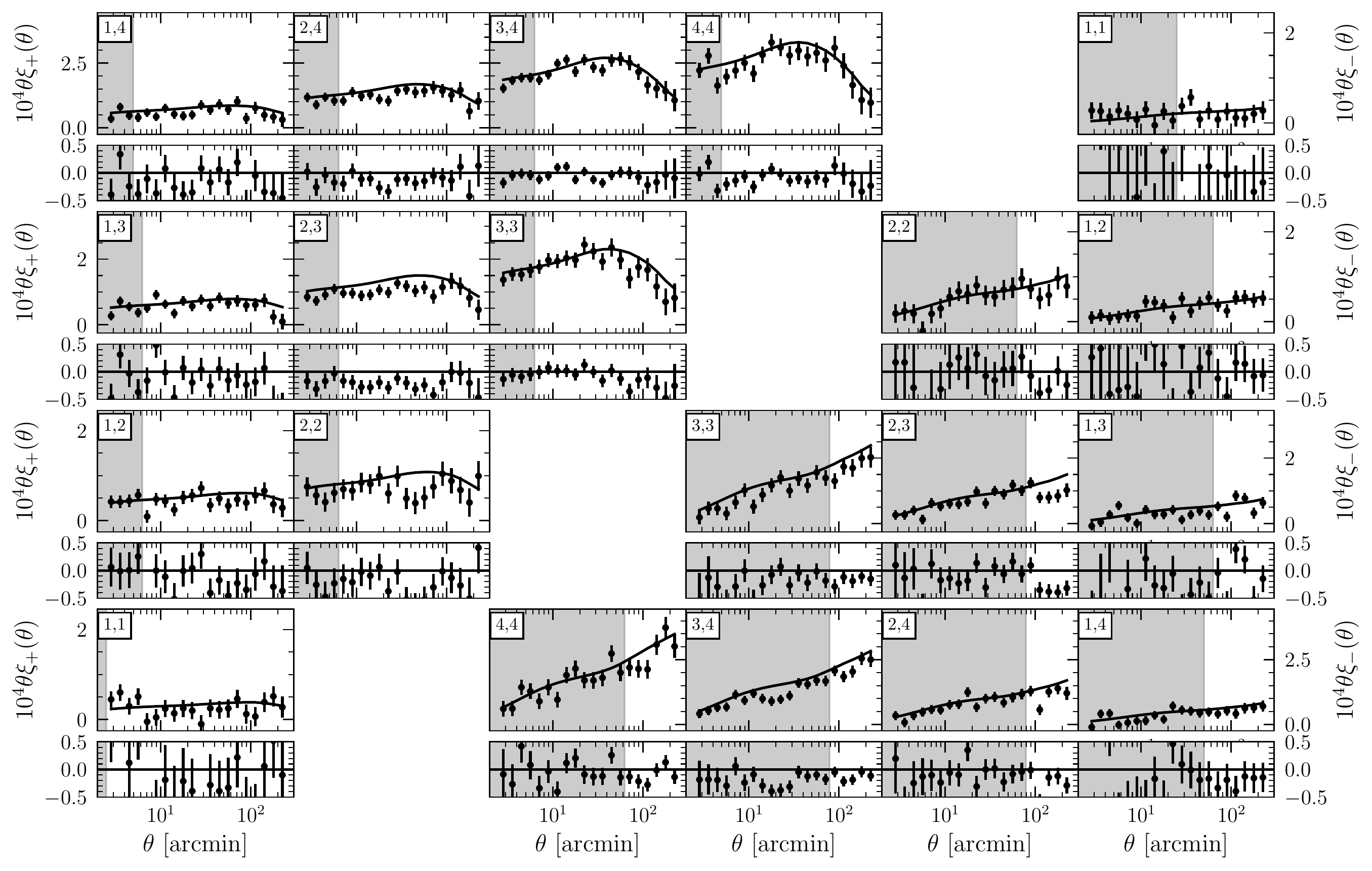}
    \caption{Comparison of \xipm\ between \buzzard\ and the DES Y3 data. As in fig. \ref{fig: 2x2_data_sims}, rows alternate between \xipm\ and the fractional difference between simulations and data, while the numbers in each panel denote which source bin pairs are plotted. The intent of this comparison is to show that the cosmic shear measurements from our simulations largely agree with those in the DES Y3 data, so that analyses performed on the simulations can be trusted to have similar constraining power to those perfomed on the DES Y3 data. The residuals shown here are largely scale independent, which is likely a representation of the imperfect match in source redshift distributions between our simulations and data, along with a small difference in the best fit cosmology from the \xipm\ analysis in the DES Y3 data and that used in the \buzzard\ simulations.}
    \label{fig: xipm_data_sims}
\end{figure*}

In \cref{fig: 2x2_data_sims,fig: xipm_data_sims} we compare the mean \ttt\ data vector from the \buzzard\ simulations to the \ttt\ data vector measured from the DES Y3 data, finding agreement at the $10-20\%$ level for nearly all scales and redshift bins. The first redshift bin in \wtheta\ is the major exception; here the \buzzard\ simulations disagree with the DES Y3 measurements by a factor of approximately $1.5$ on large scales. This bin contains the intrinsically faintest lens galaxies in our sample, so it is possible that this is a sign of a breakdown in our color-dependent galaxy clustering model at fainter magnitudes than the samples used to constrain the model parameters in \citet{DeRose2021}. It is also interesting to note that the \buzzard\ predictions for \gammat\ in this lens bin match the data quite well. In light of the observed $12\%$ discrepancy between the bias inferred from \gammat\ and \wtheta\ in the DES Y3 \redmagic\ sample \citep{y3-3x2ptkp}, the fact that \buzzard\ agrees with the DES Y3 \redmagic\ \gammat\ measurement in the first lens bin implies that we should see a $24\%$ discrepancy in \wtheta\ on large scales under the assumption of LCDM and linear bias, which is approximately what is observed.

We also see that there is a deficit of power in the \buzzard\ measurements on small scales in \gammat\ for bin combinations that include the first lens bin. This deficit is likely a result of resolution effects in our $N$-body simulations on one-halo scales, as these measurements probe the smallest physical scales shown in these figures due to the non-local nature of \gammat. We note that for \xipm\ there are no large deviations from our simulation measurements on scales smaller than the DES Y3 scale cuts, as one might expect from baryonic effects on the matter power spectrum, which have been shown to affect the matter power spectrum at the $10-30\%$ level \citep{vandaalen2019}. 

In order to confront our two-point function measurements with theoretical predictions, we require estimates of the redshift distributions of our source and lens galaxies for each tomographic bin. We have made validation of these algorithms on the simulations presented here a focus for the DES Y3 \ttt\ analysis. Each component of our redshift estimation framework including photometric redshift estimation and tomographic binning with \texttt{SOMPZ} \citep{y3-sompzbuzzard,y3-sompz}, redshift distribution uncertainty propagation with \sdir \citep{Sanchez2020,y3-sompz}, and complementary redshift distribution information from clustering cross-correlations \citep{y3-sourcewz, y3-lenswz} and galaxy--galaxy lensing ratios \citep{y3-shearratio}, has been tested extensively on the \buzzard\ simulations. Here we briefly summarize how redshift distributions are obtained from our simulations, relegating a more extensive discussion to \cref{sec:photoz} and the papers where each individual redshift estimation component are validated and applied to the DES Y3 data set.

Our lens galaxy photometric redshift estimation is relatively unchanged from that used in DES Y1. We briefly describe it here, and refer the reader to \citet{Rozo2015}, \citet{y3-lenswz} and \citet{y3-galaxyclustering} for more details. As the \redmagic\ sample is a set of bright galaxies for which we have abundant spectroscopy, we place significant confidence in the \redmagic\ in the photometric redshift estimates provided by the algorithm itself, $p(z_{\redmagic})$. These are obtained by constructing a red-sequence spectral template from a combination of spectroscopy and galaxy cluster members. In our simulations, we assume that we have a sparse, but unbiased spectroscopic training set of similar size to that used in the data. We bin lens galaxies into five tomographic bins with edges, $\{0.15, 0.35, 0.5, 0.65, 0.8, 0.9\}$, using the mean of $p(z_{\redmagic})$. To estimate the $n(z)$ for each tomographic bin, we stack four Monte-Carlo samples drawn from the $p(z_{\redmagic})$ for each galaxy.

Source galaxy tomographic binning and mean redshift distribution estimation is performed by a self-organizing-map-based algorithm called \texttt{SOMPZ}. \texttt{SOMPZ} leverages information from many-band photometry taken in small deep fields in the DES Y3 footprint in combination with secure spectroscopic and multi-band photometric redshifts from PAU+COSMOS \citep{Alarcon2020} to estimate the redshift distributions of wide field weak lensing source galaxies, accounting for selection and noise biases using the \texttt{Balrog} image simulation framework. All of these pieces of information are included in our \buzzard\ simulations, with the simplifying assumption that the redshift catalog that is used by \texttt{SOMPZ} is sparse but unbiased. We use \sdir\ to generate samples of $n(z)$ for each source tomographic bin, propagating sample variance and shot noise errors. Clustering cross correlations between lens and source galaxies are then used to select 1000 $n(z)$ samples, and the distribution of the means of these samples is used as a prior for the analyses presented in \cref{sec:config_d}.

\begin{figure*}
	\includegraphics[width=\linewidth]{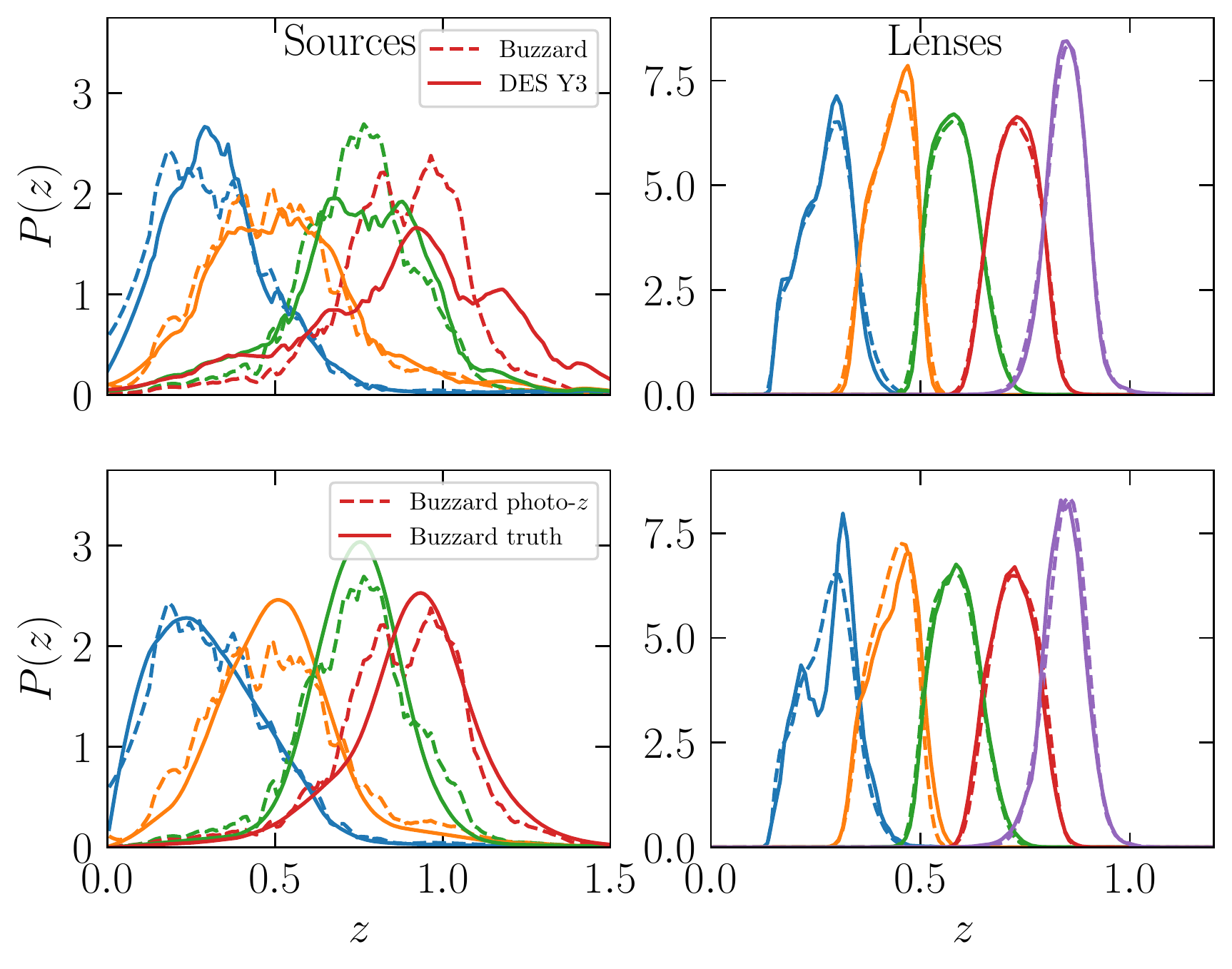}
    \caption{\textit{Top}: Comparison of source (left) and lens (right) redshift distributions in Buzzard (dashed) to DES Y3 data (solid). The mean redshifts of the source tomographic bins are  $[0.326, 0.511, 0.744, 0.871]$ in \buzzard\ compared with $[0.382, 0.563, 0.759, 0.913]$ in the Y3 data. We have matched the effective source number density and shape noise in \buzzard to that found in the DES Y3 \metacal sample \citep{y3-shapecatalog}. In combination with the close match in mean redshifts of each tomographic bin, this means that the total signal to noise of our lensing measurements in \buzzard\ should be approximately the same as that found in the DES Y3 data. \textit{Bottom}: Comparison between true (solid) and photometric (dashed) redshift distributions in Buzzard for sources (left) and lenses (right). The differences between true and photometric redshift distributions illustrated here are shown to be negligible for the simulated analyses presented in this work in \cref{sec:config_d}.}
    \label{fig:nofz}
\end{figure*}

The top row of \cref{fig:nofz} compares the source and lens redshift distributions measured from a single realization of the \buzzard\ simulations to those measured in the DES Y3 data. The agreement between the source redshift distributions in \buzzard\ and the DES Y3 data is quite good; the mean redshifts of each tomographic bin in \buzzard\ are $[0.326, 0.511, 0.744, 0.871]$ compared with $[0.382, 0.563, 0.759, 0.913]$ in the Y3 data. The high- and low-redshift tails are also captured well in the simulations, except in the last source bin where there is a tail to high redshift in the DES Y3 data that is not as extended in the \buzzard\ simulations. This discrepancy is consistent with the observation that colors in \buzzard\ approximate DES data more poorly at $z>1$, as seen in \cref{fig:colors}. A key component necessary for obtaining this level of agreement in redshift distributions was the implementation of a more realistic photometric error model, described in \cref{app:errormodel}. The agreement between the lens redshift distributions in our simulations and those found in the data is near perfect. Tuning of the red-sequence color model using the algorithm described in \cref{sec:buzzsum} and \cref{app:redsequence} was necessary in order to obtain this level of agreement.

\section{\ttt\ Likelihood} 
\label{sec:likelihood}
In this section we describe the theoretical model that that we employ in the simulated analysis of the \buzzardtwo\ \ttt\ measurements.

\subsection{Theoretical Model}
\label{sec:model}
Here we provide an overview of the model that we will use to describe the \ttt\ measurements and refer the reader to \citet{y3-generalmethods,y3-cosmicshear1,y3-cosmicshear2,y3-gglensing, y3-2x2ptbiasmodelling,y3-2x2ptmagnification} for complete technical specifications of the model. 

\subsubsection{Field-level description}
There are two main fields that we must accurately describe in order to model \ttt\ measurements: the scalar projected galaxy overdensity field measured in a tomographic bin $i$ at position $\hat{\mathbf n}$, $\delta_{\mathrm{obs}}^{i}(\hat{\mathbf n})$, and the spin-two galaxy shape field in tomographic bin $j$ at position $\hat{\mathbf n}$, $\gamma_{\alpha}^{j}(\hat{\mathbf n})$. 

We account for three contributions to the galaxy overdensity field:

\beq
\delta_{\mathrm{obs}}^{i}(\hat{\mathbf n}) = \underbrace{\int d\chi\, W^i_{\delta}\left(\chi \right)\delta_{\mathrm{g}}^{(3D)}\left(\hat{\mathbf n} \chi, \chi\right)}_{\delta_{\mathrm{D}}^{i}(\hat{\mathbf n})} + \delta_{\mathrm{RSD}}^{i}(\hat{\mathbf n}) + \delta_{\mu}^{i}(\hat{\mathbf n})\,,
\eeq
where $\chi$ is the comoving distance, and $W_{\delta}^i = n^i(z)\, d z/d\chi$ the normalized radial window function for galaxies in tomographic bin $i$. The first term in this sum, $\delta_{\mathrm{D}}^{i}(\hat{\mathbf n})$, is the projection of the three-dimensional density contrast, $\delta_{\mathrm{g}}^{(3D)}$, and the following two terms are contributions from redshift-space distortions (RSD) and magnification, respectively. 

The configuration-space three-dimensional density contrast can be perturbatively expanded in terms of operators that satisfy the symmetries of general relativity. We account for the following dependencies in this work: 

\begin{multline}
    \delta_{\mathrm{g}}^{3D} \sim f(\delta_{\mathrm{m}}, \nabla_i \nabla_j \Phi, \nabla_i v_j) \sim f^{(1)}(\delta_{\mathrm{m}}) + f^{(2)}(\delta^2_{\mathrm{m}}, s^2)  \\ + f^{(3)}(\delta^3_{\mathrm{m}},\delta_{\mathrm{m}} s^2, \psi, st) \,,\label{eq:biasexpansion}
\end{multline}
where $\delta_m$, $\Phi$. and $v$ are the three-dimensional matter density contrast, potential and velocity fields, and $\psi$, $s^2$ and $t$ are the scalar quantities constructed from contractions of the shear and velocity divergences $\nabla_i \nabla_j \Phi$ and $\nabla_i v_j$ \cite{McDonald&Roy2009}. On the right hand side of \cref{eq:biasexpansion} we have organized these fields by the order at which they contribute to the galaxy density contrast, where all arguments of the function $f^{i}$ contribute at $i$-th order, respectively. If we remain agnostic to the absolute magnitude of the contribution from these fields to the total galaxy density contrast, and treat them to one-loop order in perturbation theory, then we must introduce four bias coefficients per lens bin: $b_1$, $b_2$, $b_{3nl}$ and $b_{s^2}$ \cite{McDonald&Roy2009, Saito2014}. The parameters $b_1$ and $b_2$ describe the dependence of the galaxy density contrast on the matter density contrast and its square, $b_{3nl}$ governs the dependence of all third order fields that contribute at one-loop, and $b_{s^2}$ dictates the dependence on $s^2$. \citet{Pandey2020a} showed that for the accuracy of DES Y3, it is sufficient to leave only $b_1$ and $b_2$ entirely free, and to relate $b_{3nl}$ and $b_{s^2}$ to $b_1$ through the so-called Lagrangian co-evolution relations \cite{Matsubara2013}:

\begin{align}
    b_{s^2} &= (-4/7)\times(b_1 - 1) \\
    b_{3nl} &= b_1 - 1\,.
\end{align}
Additionally, we multiply $b_1$ by the fully non-linear matter density contrast, whose power spectrum is modeled using \texttt{halofit} \cite{Smith2003,Takahashi2012}.  In Sec. \ref{sec:validation} we explore a fiducial model that assumes only linear bias, as well as a higher-order bias model that leaves $b_1^{i}$ and $b_2^{i}$ free per lens bin, where the superscript enumerates the lens redshift bin that each coefficient contributes to. 

RSD contributes to this projected field through the apparent bulk motion of galaxies across redshift bins due to the large-scale coherent infall of galaxies towards each other \citep{Kaiser}. As our redshift bins are broad, this effect can be treated at linear order, and thus depends only on the linear matter density field, and the Hubble parameter, $H(z)$. As such, inclusion of RSD contributes no additional free parameters to our model. During the preparation of this work, a bug in the implementation of RSD in the \buzzard\ simulations was identified. As our analysis was quite advanced, we opted to correct for this by adjusting our model to account for it, rather than correcting our simulations. Additional details are contained in \cref{sec:rsd_bug}.

The magnification contribution to $\delta_{\mathrm{obs}}^{i}$ is sourced by gravitational lensing of galaxies by matter along the line-of-sight and contributes to the density both in a purely geometric manner by increasing/decreasing the apparent surface area around dense/underdense lines of sight, and by modulating the galaxy selection function through the (de)magnification of galaxy magnitudes and sizes around (under)dense lines of sight. The first of these effects depends only on the underlying cosmological model and matter power spectrum, but the second depends on the number density of the galaxy sample in question as a function of flux and size. To account for this dependence we introduce the proportionality constant $C^i$ and write the magnification term as
 \beq
  \delta_{\mu}^{i}(\hat{\mathbf n}) = C^i \kappa^i(\hat{\mathbf n})\, ,
 \eeq
where $C^i=2[\alpha^{i}(m)-1]$ and $\alpha$ is the slope of the intrinsic magnitude--size distribution, $n^\text{intr}(m)$:
\begin{equation}\label{eq:alpha}
   \alpha = 2.5\frac{d}{dm}[\log{n^\text{intr}(m)}].
\end{equation}
We have also introduced the tomographic convergence field
 \beq
 \kappa^i(\hat{\mathbf n})=\int d\chi \,W_{\kappa}^i(\chi)\delta_\mathrm{m}\left(\hat{\mathbf n} \chi, \chi\right)
 \eeq
 where $\delta_\mathrm{m}$ is the 3D matter density contrast, and the tomographic lensing efficiency
 \beq
 W_{\kappa}^i (\chi)= \frac{3\Omega_m H_0^2}{2}\int_\chi^{\chi_H} d\chi' n^i(\chi')
 \frac{\chi}{a(\chi)}\frac{\chi'-\chi}{\chi'}\,.
 \eeq
 
It is important to note that we assume constant galaxy bias, magnification, and enclosed mass (see Sec. \ref{sec:covariance}) parameters in each lens bin. The validity of this assumption is tested in our simulated analyses.

We can write the galaxy shape field as
 \beq\label{eq:shape_field}
 \gamma_{\alpha}^{j}(\hat{\mathbf n}) = \gamma_{\mathrm G,\alpha}^{j}(\hat{\mathbf n})+\epsilon_{\mathrm{I},\alpha}^{j}(\hat{\mathbf n})+\epsilon_{0,\alpha}^j(\hat{\mathbf n})\,,
 \eeq
where $\gamma_{\mathrm G,\alpha}^{j}$ is the contribution sourced by the gravitational field, i.e. the gravitational lensing signal, and the latter two terms are sourced by the intrinsic shapes of galaxies, where we have decomposed this contribution into a spatially coherent alignment with the large-scale matter distribution $\epsilon_{\mathrm{I},\alpha}^{j}(\hat{\mathbf n})$ and a stochastic contribution $\epsilon_{0,\alpha}^j(\hat{\mathbf n})$ which will only contribute to the noise in our correlation function measurements. 

In order to model the spatially coherent "intrinsic alignment" (IA) term, $\epsilon_{\mathrm{I},\alpha}^{j}(\hat{\mathbf n})$, we first construct a three-dimensional shape field $\bar{\gamma}^{\rm IA}_{ij}$ using the tidal alignment and tidal torquing (TATT) model as follows \citep{y3-cosmicshear2}:

\begin{align}\label{eq:field_IA}
\bar{\gamma}^{\rm IA}_{ij} = A_1 s_{ij} +A_{1\delta} \delta s_{ij} + A_2 s_{ik}s_{kj} + \cdots\, ,
\end{align}

\noindent where $s_{ij}$ is the traceless tidal field tensor. This three dimensional field must then be projected to obtain the two-dimensional field described in \cref{eq:shape_field}, which is then used to compute angular correlation functions.

We let the coefficients in Eq. \ref{eq:field_IA} evolve as a power law in redshift:
\begin{equation}\label{eq:theory:tatt_c1}
    A_1(z) = -a_1 \bar{C}_{1} \frac{\rho_{\rm crit}\Omega_{\rm m}}{D(z)} \left(\frac{1+z}{1+z_{0}}\right)^{\eta_1},
\end{equation}
\begin{equation}\label{eq:theory:tatt_c2}
    A_2(z) = 5 a_2 \bar{C}_{1} \frac{\rho_{\rm crit}\Omega_{\rm m}}{D^2(z)} \left(\frac{1+z}{1+z_{0}}\right)^{\eta_2},
\end{equation}
\noindent
$\bar{C}_1$ is a normalisation constant, by convention fixed at a value $\bar{C}_1=5\times10^{-14}M_\odot h^{-2} \mathrm{Mpc}^2$, obtained from SuperCOSMOS (see \citealt{brown02}). The parameter $z_{0}$ is a pivot redshift, which we fix to the value 0.62, $\rho_{\rm crit}$ is the critical density, and $D(z)$ is the linear growth function.

We then relate $A_{1\delta}$ to $A_1$ via

\begin{equation}
A_{1\delta} = b_{\mathrm{TA}} A_1,
\end{equation}
\noindent where $b_{\mathrm{TA}}$ can be interpreted as the source galaxy bias, although we have allowed it to assume a broader prior than would be physically possible if it were a galaxy bias parameter in order to give this term more flexibility in our intrinsic alignment (IA) model. In total, we have five IA parameters: three amplitude parameters in $a_1$, $a_2$, $b_{\rm TA}$, and two parameters governing their redshift evolution, $\alpha_1$ and $\alpha_2$. This three-dimensional field can be projected into the two-dimensional shape field that is observed. 

Note that we include modeling of IAs in our simulated analyses solely for sake of comparison with the expected constraining power of the analyses being performed on the data. No IA signal is included in the \buzzard\ simulations. Marginalizing over the full IA model significantly impacts the constraining power of cosmic shear analyses, but it's impact on \ttt\ is much less severe \citep{y3-cosmicshear2, y3-3x2ptkp}. Nevertheless, we have verified that none of the results presented here change in a qualitative way when we do not marginalize over IAs.

\subsubsection{Angular Two-point Statistics}
We are not interested in these fields themselves, but rather in their angular two-point auto- and cross-correlations. In general, an angular two-point function $\xi^{ij}_{AB}(\theta)$, where $i$ and $j$ are tomographic bin indices, and $A$ and $B$ specify the fields being correlated, can be related to its corresponding angular power spectrum $C^{ij}_{AB}(\ell)$ via the relation
\beq
\xi^{ij}_{AB}(\theta) = \sum_{\ell} \frac{2\ell + 1}{4\pi}C^{ij}_{AB}(\ell)d_{mn}(\theta)
\eeq
where $d_{mn}$ is the Wigner D-matrix, and $m=n=0$ for $w(\theta)$, $m=0$, $n=2$ for $\gamma_t$ and $m=2$, $n=\pm 2$ for $\xi_{\pm}$. 

We compute bin-averaged predictions for these two-point functions by computing $d_{mn}(\theta)$ averaged over the width of the angular bin as described in \citet{y3-generalmethods}. The angular power spectra are in turn computed by projecting three-dimensional power spectra $P_{AB}$ along the line of sight weighted by the relevant projection kernels. For $\xi_{\pm}$ and $\gamma_t$ we use the Limber approximation \citep{Limber1953}
 
\begin{align}
 C_{AB}^{ij}(\ell) = \int d\chi \frac{W_A^i(\chi)W_B^j(\chi)}{\chi^2}P_{AB}\left(k = \frac{\ell+0.5}{\chi},z(\chi)\right)\,,
\end{align}
but for $w(\theta)$, the accuracy of the DES Y3 data demands that we compute the full non-Limber projection integral on large scales as described in \citet{y3-generalmethods,Fang2020}. Our simulations do not assume the Limber approximation for \wtheta, but our ray-tracing algorithm does implicitly assume it and thus the \gammat\ and \xipm\ measurements in our simulations also use the Limber approximation. Additionally, whereever the non-linear matter power spectrum appears in our analysis, we use the re-calibrated \Halofit\ model \citep{Smith2003}, as described in \citet{Takahashi2012}.

We also wish to mitigate the non-local nature of the galaxy--galaxy lensing signal, as \gammat\ at a fixed value of $\theta$ is sensitive to the total mass enclosed at all angles less than $\theta$. There are a number of similar methods that aim to remove the sensitivity of \gammat\ to this enclosed mass \citep{Baldauf2012, MacCrann2019, Park2020}, but we opt to analytically marginalize over a single parameter in each lens bin in order to account for this:
\beq
\gamma^{ij}_{t}(\theta) = \gamma^{ij}_{t,\rm model}(\theta) + \frac{C_{ij}}{\theta^2}\, ,
\eeq
where 
\begin{align}
C_{ij} &= B_{i}\int dz_{l}dz_{s}n^{i}(z_{l})n^{j}(z_{s})\Sigma_{\rm crit}^{-1}(z_{l},z_s)D^{-2}_{A}(z_l) \label{eq:thinlens}\\
&\equiv B_{i}\beta_{ij}
\end{align}
where $B_{i}$ is a free parameter describing the mass enclosed within the minimum scale used in the \gammat\ part of our data vector per lens bin, $\Sigma_{\rm crit}$ is the critical surface mass density, and $D_{A}(z_{l})$ is the angular diameter distance to $z_l$. Notably, we have made the assumption that this enclosed mass does not evolve significantly over the width of our lens redshift bin, an assumption that is implicitly tested in this work.

Instead of sampling over these additional enclosed mass parameters, analytically marginalize over them by adding an extra component to our covariance matrix as described in \cref{sec:covariance}.

\subsection{Scale cuts}
\label{sec:scalecuts}
We have determined a set of scale cuts such that our fiducial linear bias modeling choices detailed above are sufficient to deliver unbiased cosmological constraints in a series of simulated analyses on noiseless data vectors. When making choices about whether a particular analysis assumption is acceptable for the constraining power afforded us by the DES Y3 data, we have set a criterion that the assumption under consideration must result in no greater than a $0.3\sigma$ bias in the two-dimensional spaces of $S_8-\Omega_{m}$ and $w-\om$ for $\Lambda$CDM and $w$CDM analyses respectively. Formally, our criterion can be expressed as:
\beq\label{eq:noiseless_scalecut}
P(\hat{S}_8,\hat{\Omega}_m | \mathbf{d}_{\rm fid}) > 0.235\, ,
\eeq
where $(\hat{S}_8, \hat{\Omega}_m) = \mathrm{E}[P(S_8,\Omega_m | \mathbf{d}_{\rm cont})]$ for the $\Lambda$CDM case and likewise for $w-\om$ for the $w$CDM case. Procedurally, this test is performed as follows. We generate a data vector, $\mathbf{d}_{\rm cont}$ by breaking a set of assumptions that are made in our fiducial analysis. We then analyze $\mathbf{d}_{\rm cont}$ using the fiducial model where those assumptions hold, producing the posterior $P(S_8,\Omega_m | \mathbf{d}_{\rm cont})$. We then generate a data vector, $\mathbf{d}_{\rm fid}$, using the same cosmology as that assumed when generating $\mathbf{d}_{\rm cont}$, but now making the assumptions that were previously broken. We analyze this with the same model, giving us the posterior $P(S_8,\Omega_m | \mathbf{d}_{\rm fid})$. Our scale cut criterion, Eq. \ref{eq:noiseless_scalecut}, then requires that $\mathrm{E}[P(S_8,\Omega_m | \mathbf{d}_{\rm cont})]$ must fall within the $0.3\sigma$ ($P>0.235$) confidence region of $P(S_8,\Omega_m | \mathbf{d}_{\rm fid})$.

The contaminated data vector that we use, $\mathbf{d}_{\rm cont}$, breaks our matter power spectrum and linear bias assumptions. Instead of our fiducial assumptions, we use a model for the non-linear matter power spectrum that takes into account baryonic effects as measured in the OWLS AGN simulation \citep{vandaalen2011}. OWLs AGN possesses feedback effects that are more significant than many more recent hydrodynamic simulations, but are still within the realm of possibility. Additionally, we contaminate with the non-linear bias model described in Sec \ref{sec:model}, and bias coefficients as described in \citet{y3-2x2ptbiasmodelling}.

We have determined that scale cuts on \xipm\ that yield a $\chi^2$ difference between our contaminated and fiducial models for \xipm\ of $0.5$ are sufficient to pass Eq. \ref{eq:noiseless_scalecut}. For $\gammat$ and $\wtheta$, we make angular scale cuts that correspond to $6\, \mpc$ and $8\, \mpc$ (or $4\, \mpc$ for both $\gammat$ and $\wtheta$ when testing non-linear bias modeling) respectively at the low edge of the redshift range for each lens bin. This procedure, along with a number of additional stress tests of our fiducial model, including tests of different matter power spectra, higher-order lensing effects, and a more complex IA redshift scaling, are discussed further in \citet{y3-generalmethods}.

\section{Validation of the DES Y3 \ttt\ Analysis}
\label{sec:validation}
We now proceed to investigate whether the models described in Sec. \ref{sec:likelihood} are sufficient to recover unbiased constraints on cosmological parameters in our simulations. We investigate four different analysis configurations:
\begin{enumerate}[label=\Alph*.]
    \item Fixed cosmology;
    \item Linear bias, true redshift distributions;
    \item Non-linear bias, true redshift distributions; and
    \item Linear bias, calibrated \photoz\ distributions.
\end{enumerate}
The results of these tests are described in the following subsections. The parameters that we leave free for each of these configurations are listed with their priors in Table \ref{tab:params}. We also list the part of the data vector that each parameter contributes to, i.e. \xipm, \wtheta, or \gammat. 

The simulations presented in this work contain complexities beyond the assumptions made in our fiducial model, including
\begin{enumerate}
    \item beyond one-loop galaxy bias, including stochasticity;
    \item redshift evolution of galaxy bias within each lens bin;
    \item redshift evolution of enclosed mass parameter, $B_i$, within each lens bin;
    \item source galaxy clustering;
    \item reduced shear;
    \item source galaxy magnification;
    \item multiple lens plane deflection (``beyond Born'' approximation);
    \item anisotropic source and lens $n(z)$; and
    \item non-Gaussian distributed data vectors.
\end{enumerate}

For all posterior parameter distributions presented in this paper, we make use of the \texttt{PolyChord} nested sampler \citep{Handley2015} with the same settings used in \citet{y3-3x2ptkp}, which have been shown to yield converged posteriors and evidences. All fits are done on data vectors with the fiducial scale cuts described in Section \ref{sec:scalecuts} unless explicitly stated otherwise. We make use of the \cosmosis\footnote{https://bitbucket.org/joezuntz/cosmosis/} likelihood and sampling framework \citep{Zuntz2015}, which is one of two essentially interchangeable implementations of the model described in \cref{sec:model}, along with \COSMOLIKE \citep{Krause&Eifler2016}. \cosmosis makes use of the \texttt{CAMB} Boltzmann solver \citet{Howlett2012, Lewis2000}.

We make use of the covariance matrix appropriate for a single survey realization even though we are fitting to the mean of 18 simulations. This is because we wish to keep as close to the actual analysis that will be performed on the DES Y3 data as possible. Analyzing each of the 18 simulations independently with this covariance matrix and taking the product of their posteriors would have yielded the closest approximation to the analyses performed on the DES Y3 data, but doing so would have proven extremely computationally expensive. Instead, we have opted to run analyses on the mean data vector from all 18 simulations, but use a covariance matrix appropriate for a single realization. This is in the spirit of the rest of the simulated analyses used to validate the DES Y3 cosmological constraints. 

Because of this choice, we are susceptible to the same parameter projection effects in marginalized posteriors as those discussed in \citet{y3-generalmethods}. These parameter projection effects are a consequence of correlations between cosmological parameters, and poorly constrained nuisance parameters, such that when the nuisance parameters are marginalized over they impart a bias in the marginalized posterior mean of a parameter of interest. We emphasize that these effects are not a systematic error, in the sense that they decrease in size as the constraining power of an analysis increases. For a concrete example of parameter projection effects, see the discussion in \cref{sec:config_b}. 

As such, we always compare posteriors obtained from our \buzzard\ simulated data vectors with posteriors obtained by running the same analysis on a synthetic data vector generated by \cosmosis\ assuming the \buzzard\ cosmology, true redshift distributions, and best-fit nuisance parameters from analysis configuration A1 or A2 depending on whether the analysis under consideration uses a linear or non-linear bias model. We will often refer to the \cosmosis\ simulated data vector as an \textit{uncontaminated} data vector, as it is produced from the same model that is being used to perform the analysis. When comparing the posteriors obtained from the \buzzard\ simulated data vector to those obtained from the uncontaminated data vector, we can disentangle the parameter biases resulting from parameter projection effects that will be present in \textit{both} marginalized posteriors from parameter biases resulting from unmodeled systematics in \buzzard\, which will only be present in the \buzzard\ posteriors. 

In order to illustrate the increased constraining power available to us through the use of the mean measurement from 18 simulations, we show $1/\sqrt{18}\sigma$ and $2/\sqrt{18}\sigma$ confidence intervals when plotting constraints run on the \buzzard\ data vectors. These roughly correspond to the $1\sigma$ and $2\sigma$ confidence intervals that we would obtain, had we performed analyses using the single Y3 covariance scaled by the inverse of the number of simulations. In fact, these sets of confidence intervals are identical in the limit of flat priors, which we use for the majority of our parameters. If our inference framework worked perfectly, we would expect the mean of the posteriors derived from 18 simulations to match the mean of the posteriors from the uncontaminated analysis at a level defined by these tighter confidence regions, even though the uncertainties from a Y3--sized data set are a factor of $\sqrt{18}$ larger.

Because we have a finite number of simulations, we need to apply a slightly different criterion than Eq. \ref{eq:noiseless_scalecut} in order to validate our analysis assumptions. In particular, considering the data vector generated by taking the mean over all of our simulations as $\mathbf{d}_{\rm cont}$ in Eq. \ref{eq:noiseless_scalecut}, we note that the quantity $E[P(S_8,\Omega_m | \mathbf{d}_{\rm cont})]$ now has an uncertainty associated with it due to the noise that remains in $\mathbf{d}_{\rm cont}$ even after averaging over 18 simulations. We wish to take this uncertainty into account when determining whether an analysis passes or fails our criterion. In order to do this, we can generalize Eq. \ref{eq:noiseless_scalecut} to:
\begin{align}
\textrm{PTE} &\equiv & 1 - \int_{P(S_8,\Omega_m |\mathbf{d}_{\rm fid})>0.235} P(S_8,\Omega_m | \mathbf{d}_{\rm cont};\Sigma/N)d\theta & \nonumber \\
& \approx &  1 - \int_{P(S_8,\Omega_m |\mathbf{d}_{\rm fid})>0.235} P(S_8,\Omega_m | \mathbf{d}_{\rm cont};\Sigma)^{N}d\theta \, ,
\label{eq:noisy_scalecut}
\end{align}
where $\Sigma$ is the covariance for a single Y3 \buzzard\ realization and $N$ is the number of simulation realizations (18). The proportionality in the second line holds in the limit that the prior in the two-dimensional space of $S_8-\Omega_m$ is flat over the region where $P(S_8,\Omega_m |\mathbf{d}_{\rm fid})>0.235$. This quantity is the probability that the analysis performed on the mean buzzard data vector, $\textbf{d}_{\rm cont}$, results in a parameter bias that is more than $0.3\sigma$, i.e. a probability to exceed (PTE) $0.3\sigma$ cosmological parameter bias.

\textit{Values for these probabilities are quoted in \cref{tab:pte} and these constitute the main result of this paper}. In particular, we find that in all cases the probability for any analysis to exceed a $0.3\sigma$($1\sigma$) bias is less than $62\%$ ($2\%$) for $\Lambda$CDM and $58\%$ ($1\%$) in $w$CDM. In the following sub-sections, we break these results down by analysis configuration. 

We supplement these values by calculating the mean two-dimensional offsets in $\Omega_m - S_{8}$ and $S_{8} - w$ in \cref{tab:mean_bias} for \LCDM and $w$CDM, respectively. These mean biases are more easily interpretable, but we caution that these are less robust than the PTE values, as the mean posterior values for the \buzzard\ chains have non-negligible uncertainty of approximately $1/\sqrt{18}\sigma$ due to residual noise in the mean \buzzard\ data vector. That uncertainty is an upper bound on the actual uncertainty on these numbers, since our measurements do not include shape noise, while our covariance matrix does.

\begin{table*}
\caption{Parameters and priors}
\begin{center}
\begin{tabular}{| c  c  c  c |}
\hline
\hline
Parameter & Prior & Data-vector & Analysis Configuration\\  
\hline 
\multicolumn{4}{|c|}{{\bf Cosmology}} \\
$\Omega_m$  &  flat (0.1, 0.9) & \cs, \wtheta, \gammat & B,C,D \\ 
$A_s$ &  flat ($5\times 10^{-10},5\times 10^{-9}$) & \cs, \wtheta, \gammat & B,C,D  \\ 
$n_s$ &  flat ($0.87, 1.07$) &\cs, \wtheta, \gammat & B,C,D  \\
$\Omega_b$ &  flat ($0.03, 0.07$) & \cs, \wtheta, \gammat & B,C,D  \\
$h$  &  flat ($0.55, 0.91$) & \cs, \wtheta, \gammat & B,C,D  \\
\hline
\multicolumn{4}{|c|}{{\bf Lens Galaxy Bias}} \\
$b_{1}^{i} (i=1,5)$\footnote{Analysis setup C samples over $b_1^i\sigma_8$}   & flat ($0, 3.0$) & \wtheta, \gammat & A1/2,B,C,D \\
$b_{2}^{i} (i=1,5)$\footnote{Analysis setup C samples over $b_2^i\sigma_8^2$}   & flat ($-5, 5$) & \wtheta, \gammat & A2,C \\

\hline
\multicolumn{4}{|c|}{{\bf Intrinsic Alignment}} \\
$a_{1}$   & flat ($-5,5$) & \cs, \gammat & B, C, D\\
$a_{2}$   & flat ($-5,5$) & \cs, \gammat & B, C, D\\
$\alpha_1$   & flat ($-5,5$) & \cs, \gammat & B, C, D\\
$\alpha_2$   & flat ($-5,5$) & \cs, \gammat & B, C, D \\
$b_{ta}$ & flat ($0,2$)& \cs, \gammat & B, C, D\\
\hline 
\multicolumn{4}{|c|}{{\bf Magnification}} \\
$\alpha^{i}_{\rm mag}$ & flat($-4,4$) & \gammat, \wtheta & A1/2 \\
\hline
\hline 
\multicolumn{4}{|c|}{{\bf Point Mass}} \\
$B^{i}$ & flat($-50,50$) & \gammat & A1/2\footnote{Marginalized over analytically in B/C/D} \\
\hline
\multicolumn{4}{|c|}{{\bf Lens \photoz}} \\
$\Delta z^1_{\rm l}$  & Gauss ($0.000, 0.004$) & \wtheta, \gammat & D \\
$\Delta z^2_{\rm l}$  & Gauss ($0.000, 0.003$) & \wtheta, \gammat & D \\
$\Delta z^3_{\rm l}$  & Gauss ($0.000, 0.003$) & \wtheta, \gammat & D \\
$\Delta z^4_{\rm l}$  & Gauss ($0.000 , 0.005$)& \wtheta, \gammat & D \\
$\Delta z^5_{\rm l}$  & Gauss ($0.000, 0.01$) & \wtheta, \gammat & D  \\
$\sigma z^5_{\rm l}$  & Gauss ($1.000, 0.054$) & \wtheta, \gammat & D  \\

\hline
\multicolumn{4}{|c|}{{\bf Source \photoz\ }} \\
$\Delta z^1_{\rm s}$  & Gauss ($0.000, 0.018$) & \cs, \gammat & D \\
$\Delta z^2_{\rm s}$  & Gauss ($0.000, 0.013$) & \cs, \gammat & D \\
$\Delta z^3_{\rm s}$  & Gauss ($0.000, 0.006$) & \cs, \gammat & D \\
$\Delta z^4_{\rm s}$  & Gauss ($0.000 , 0.013$)& \cs, \gammat & D \\
\hline
\multicolumn{4}{|c|}{{\bf Shear calibration}} \\
$m^{i} (i=1,4)$ & Gauss ($0.0, 0.015$)& \cs, \gammat & D\\
\hline
\end{tabular}
\end{center}
\label{tab:params}
\end{table*}

\begin{table*}
\caption{Best-fit configuration A1/2 $\chi^2$/data vector dimensionality ($N_{\rm DV}$)/number of free parameters ($N_{\rm param}$)}
\begin{center}
\begin{tabular}{| c || c | c | c || c | c | c | }
\hline
\hline
&  A1 $\chi^2$ & A1 $N_{\rm DV}$ & A1 $N_{\rm param}$ &  A2 $\chi^2$ & A2 $N_{\rm DV}$ & A2 $N_{\rm param}$ \\  
\hline 
\cs & 1.4 & 207 & 0 &  & -- & -- \\
\hline 
\wtheta & 4.5 & 53 & 15 & 7.2 & 68 & 20 \\
\hline 
\gammat & 9.1 & 232 & 15 & 8.4 & 272 & 20 \\
\hline 
\end{tabular}
\end{center}
\label{tab:config_a_chisq}
\end{table*}

\begin{table*}
\caption{Probability to Exceed (PTE) $0.3/1\sigma$ Cosmological Parameter Bias. PTE values less than $1\%$ are indicated as $<$0.01.}
\begin{center}
\begin{tabular}{| c | c | c | c | c | c | c |}
\hline
\hline
& $\Lambda$CDM B & $w$CDM B & $\Lambda$CDM C & $w$CDM C & $\Lambda$CDM D & $w$CDM D\\  
\hline 
\cs                 & 0.43/$<$0.01 & 0.19/$<$0.01 & --           & --           & 0.14/$<$0.01 & 0.24/$<$0.01\\
\hline 
\wtheta,\gammat     & 0.25/$<$0.01 & 0.35/$<$0.01 & 0.20/$<$0.01 & 0.26/$<$0.01  & 0.08/$<$0.01 & 0.05/$<$0.01\\
\hline 
\cs,\wtheta,\gammat & 0.61/0.02    & 0.49/$<$0.01 & 0.35/$<$0.01 & 0.58/0.01 & 0.15/$<$0.01 & 0.12/$<$0.01\\
\hline 
\end{tabular}
\end{center}
\label{tab:pte}
\end{table*}

\begin{table*}
\caption{Mean $\Omega_m-S_{8}$ / $w-S_{8}$ Parameter Bias}
\begin{center}
\begin{tabular}{| c | c | c | c | c | c | c |}
\hline
\hline
& $\Lambda$CDM B & $w$CDM B & $\Lambda$CDM C & $w$CDM C & $\Lambda$CDM D & $w$CDM D\\
\hline 
\cs & $0.19\sigma$ & $0.04\sigma$ & -- & -- & $0.07\sigma$ & $0.15\sigma$\\
\hline 
\wtheta,\gammat & $0.13\sigma$ & $0.14\sigma$ & $0.05\sigma$ & $0.11\sigma$ & $0.01\sigma$ & $0.01\sigma$ \\
\hline 
\cs,\wtheta,\gammat & $0.23\sigma$ & $0.18\sigma$ & $0.09\sigma$ & $0.21\sigma$ & $0.07\sigma$ & $0.05\sigma$\\
\hline 
\end{tabular}
\end{center}
\label{tab:mean_bias}
\end{table*}

\subsection{Fixed cosmology}
\label{sec:config_a}
Before we obtain cosmological constraints from our simulated measurements, we first wish to demonstrate that our models for galaxy bias and lens magnification provide good fits to our measurements when the cosmology in our model is fixed to the true cosmology of our simulations. Additionally, we fix all photometric redshift and shear calibration parameters and use the measured true source and lens redshift distributions to make our model predictions. We set all IA parameters to zero as our simulations contain no IA contribution. We refer to this as analysis configuration A. 

Within this configuration, we consider two variants: one using a linear bias model, and one using our full non-linear bias model described in \cref{sec:model}, which we refer to as configuration A1 and A2, respectively. A1 assumes one linear bias coefficient $b_1^{i}$ and one magnification coefficient $C^{i}$ for each lens bin. Analysis configuration A2 also fits for one second order bias parameter per lens bin, $b_2^{i}$. In this configuration we also sample over the enclosed-mass parameters, $B^{i}$, described in \cref{sec:model}, rather than analytically marginalizing over them in order to aid in presentation of our results. Analytic marginalization results in the same best-fit likelihood values as sampling over the enclosed-mass parameters, but interpreting model residuals in a visual format is made much easier if we can obtain best-fit values for these parameters. The priors that we assume on our bias, magnification, and enclosed-mass parameters are listed in \cref{tab:params}.

\begin{figure*}
\centering
\includegraphics[width=\linewidth]{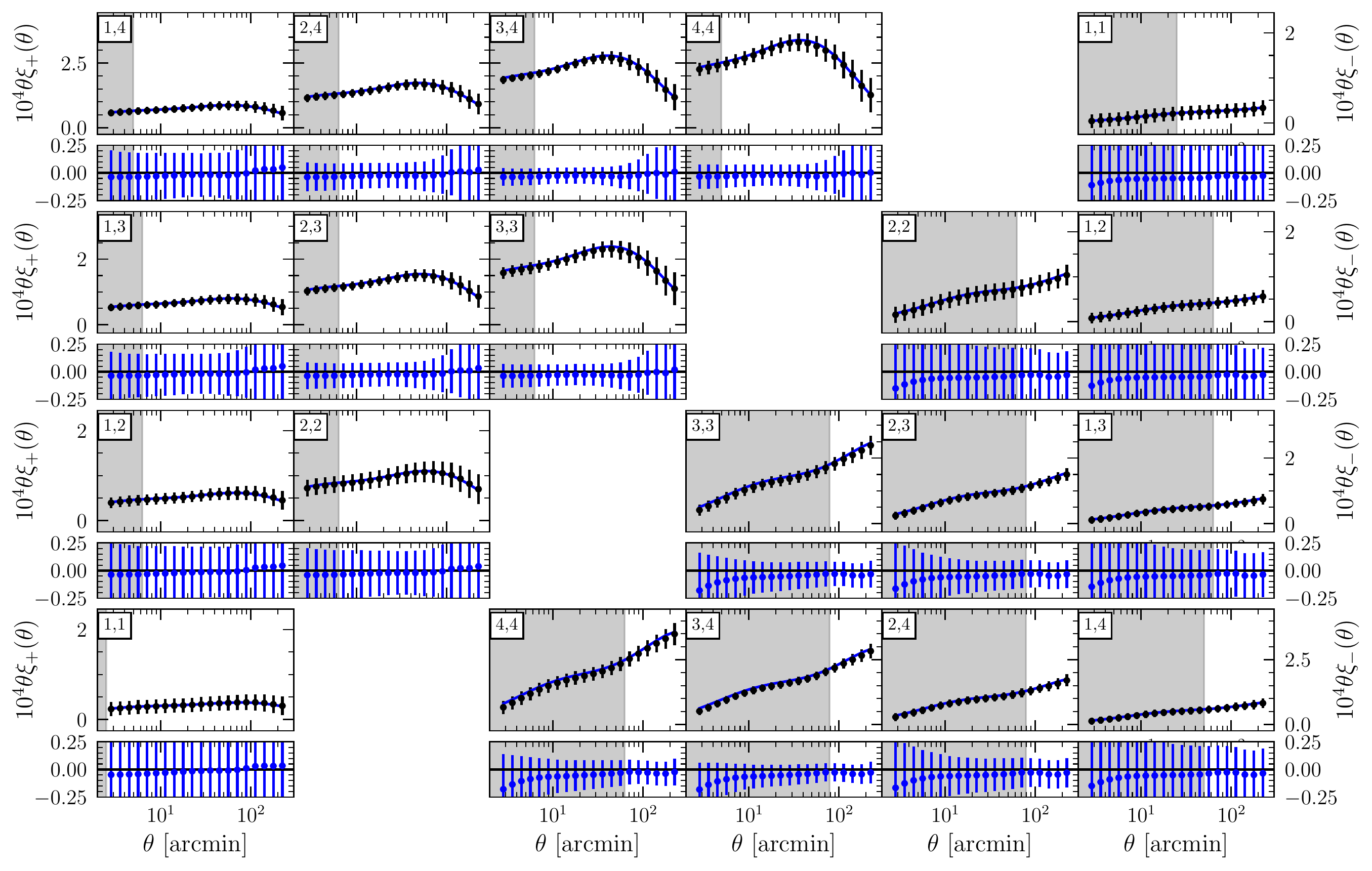}
\caption{Comparison of our fiducial \xipm\ model prediction (lines) at the true \buzzard\ cosmology to the mean measurement from our simulations without shape noise (points). Rows alternate between the signals themselves, and fractional deviations between models and simulations, and the grey regions are our fiducial scale cuts. We find $\chi^2=0.78$ for $\xi_{+}$ and $\chi^2=0.59$ for $\xi_{-}$. The differences on large scales are likely caused by a combination of source galaxy clustering and source galaxy magnification \citep{y3-generalmethods, y3-cosmicshear2}. On scales below the scale cuts used in this analysis, the observed differences are likely sourced by ray-tracing resolution effects \citep{DeRose2019}.}
\label{fig:config_a_xipm}
\end{figure*}

\begin{figure*}
    \centering
    \includegraphics[width=\linewidth]{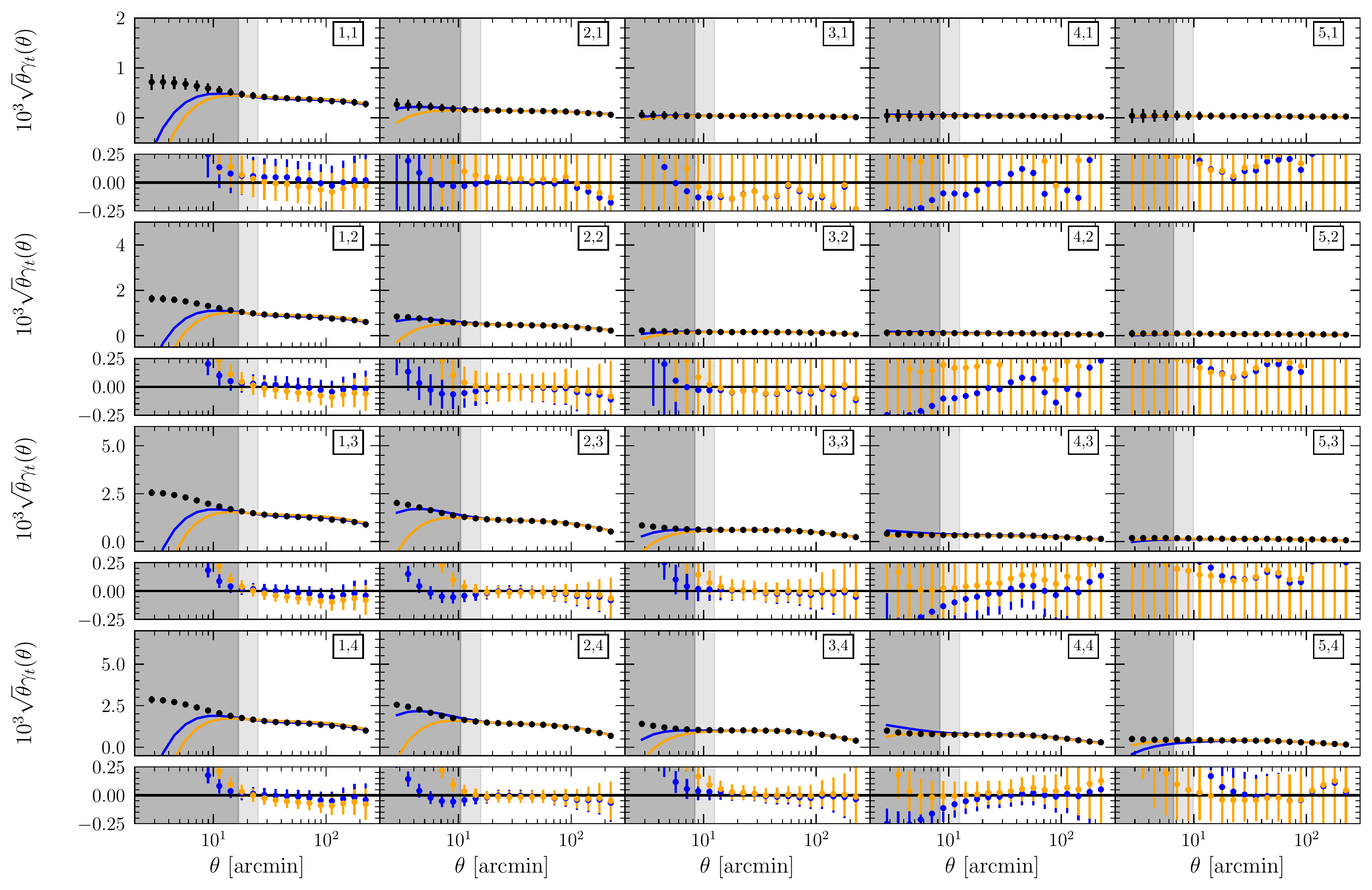}
    \includegraphics[width=\linewidth]{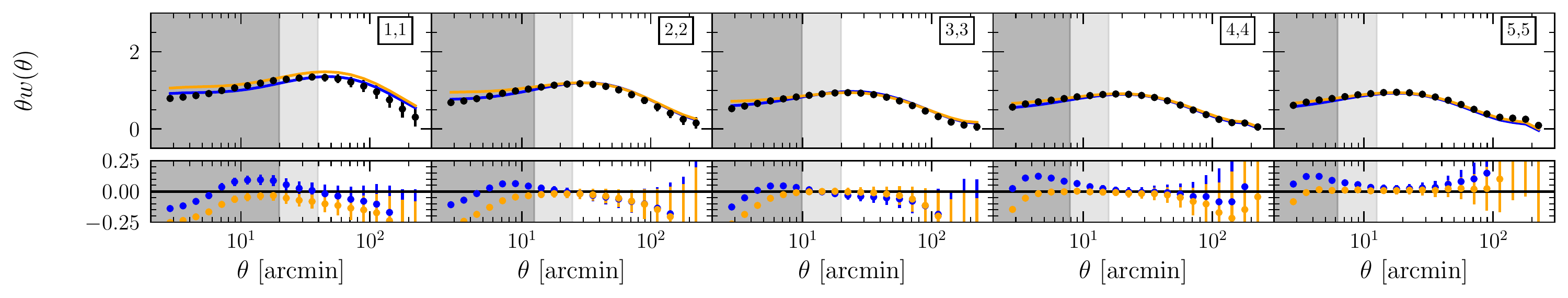}
    
    \caption{Same as Fig. \ref{fig:config_a_xipm}, but for \gammat\ and \wtheta. Best-fit configuration A1 (linear bias, blue) and A2 (non-linear bias, yellow) models fixed to the true \buzzard\ cosmology are compared to the mean \buzzard\ data vector without shape noise. Rows alternate between the signals themselves, and fractional deviations between models and simulations. For configuration A1 we find $\chi^2=4.5$ for \gammat\ and $\chi^2=9.1$ for \wtheta\ for our fiducial scale cuts, shown as light gray shaded regions. For A2 we find $\chi^2=7.2$ for \gammat\ and $\chi^2=8.4$ for \wtheta\ using $r_{\rm min}=4\mpc$ scale cuts, depicted by the dark grey shaded regions.}
    \label{fig:config_a_2x2}
\end{figure*}

The best-fit models for configuration A1 and A2 are shown in \cref{fig:config_a_xipm,fig:config_a_2x2}, and the error bars are obtained from the covariance of a single Y3 simulation, which is what is used to fit these models. The prediction for \xipm\ has no free parameters in this configuration, and we find a chi-squared of $1.4$ for $207$ data points. The differences on large scales in \xipm\ are likely caused by a combination of source galaxy clustering and source galaxy magnification \citep{y3-generalmethods, y3-cosmicshear2}. On scales below the scale cuts used in this analysis, the observed differences are likely sourced by ray-tracing resolution effects \citep{DeRose2019}.  For analysis configuration A1 and A2, the predictions for \wtheta\ and \gammat\ have 15 and 20 free parameters respectively with 53 and 232 data points for \wtheta\ and \gammat\ for A1, and 68 and 272 data points for \wtheta\ and \gammat\ for the $r_{\rm min}=4\, \mpc$ scale cuts used for the A2 analysis. A1 results in a chi-squared of 4.5 and 9.1 for \wtheta\ and \gammat, while A2 results in chi-squared values for \wtheta\ and \gammat\ of $7.2$ and $8.4$. The fact that the reduced chi-squared values for \wtheta\ and \gammat\ are larger than for \xipm\ is expected, as \wtheta\ and \gammat\ have an additional contribution from shot-noise in the lens galaxy sample, which is not present in \xipm. In all cases, the residuals are significantly smaller than the expected errors on our Y3 measurements. In the following sections, we investigate the effect of these residuals on cosmological parameter constraints, where we have formal requirements on acceptable parameter biases. We have also checked that the values for the magnification coefficients obtained in these analyses are consistent with the expected values, as further explored in \citet*{y3-2x2ptmagnification}. These results are summarized in \cref{tab:config_a_chisq}.

\begin{figure*}
	\includegraphics[width=\linewidth]{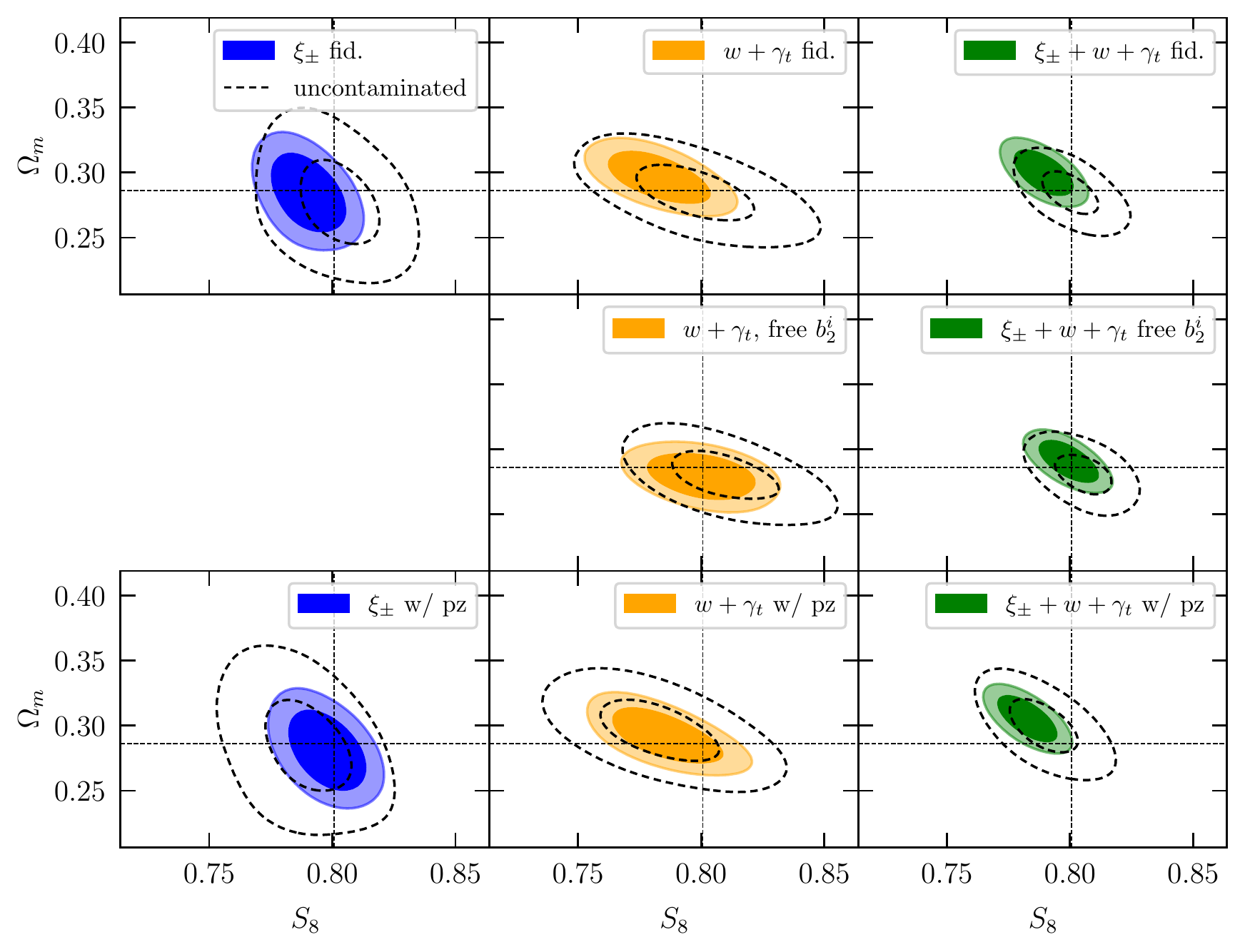}
    \caption{Constraints on \seight\ and \om\ from 1$\times$2- (\textit{left}), 2$\times$2- (\textit{middle}), and 3$\times$2- (\textit{right}) point analyses on the mean data vector from the full suite of simulations. ({\it Top}) Constraints marginalizing over only cosmology, linear bias, and IAs, assuming the true source and lens redshift distributions. Posteriors obtained using the mean \buzzard\ data vector are shown as solid lines, while dashed contours use a data vector generated at the true cosmology of the simulations with the best-fit linear bias model from analysis configuration A1, i.e., the blue line in Fig. \ref{fig:config_a_2x2}. The shaded \buzzard\ contours are the $1/\sqrt{18}$ and $2/\sqrt{18}$ confidence regions, while the dashed contours represent the 0.3 and 1 $\sigma$ confidence regions for a single simulation realization. The cross-hairs represent the true \buzzard\ cosmology and the the difference between the dashed contours and these is a product of parameter projection effects. ({\it Middle}) Same as top row, but posteriors are obtained using analysis configuration C (non-linear bias), where the uncontaminated data vector is the best fit non-linear bias model from analysis configuration A2, i.e., the yellow line in Fig. \ref{fig:config_a_2x2}. ({\it Bottom}) Same as top and middle rows, but using analysis configuration D, i.e., using calibrated photometric redshift distributions to make our model predictions, and marginalizing over source and lens photometric redshift uncertainties. This isolates the effect of photometric redshift biases on our analysis. Dashed contours are the same as the solid contours in the top row, but scaled to represent the constraining power of a single Y3 simulation.  In all cases, the probability to exceed a parameter bias of more than $0.3\sigma$ is less than $60\%$, as summarized in \cref{tab:pte} and \cref{tab:mean_bias}.}
    \label{fig:config_bcd_lcdm}
\end{figure*}

\begin{figure*}
	\includegraphics[width=\linewidth]{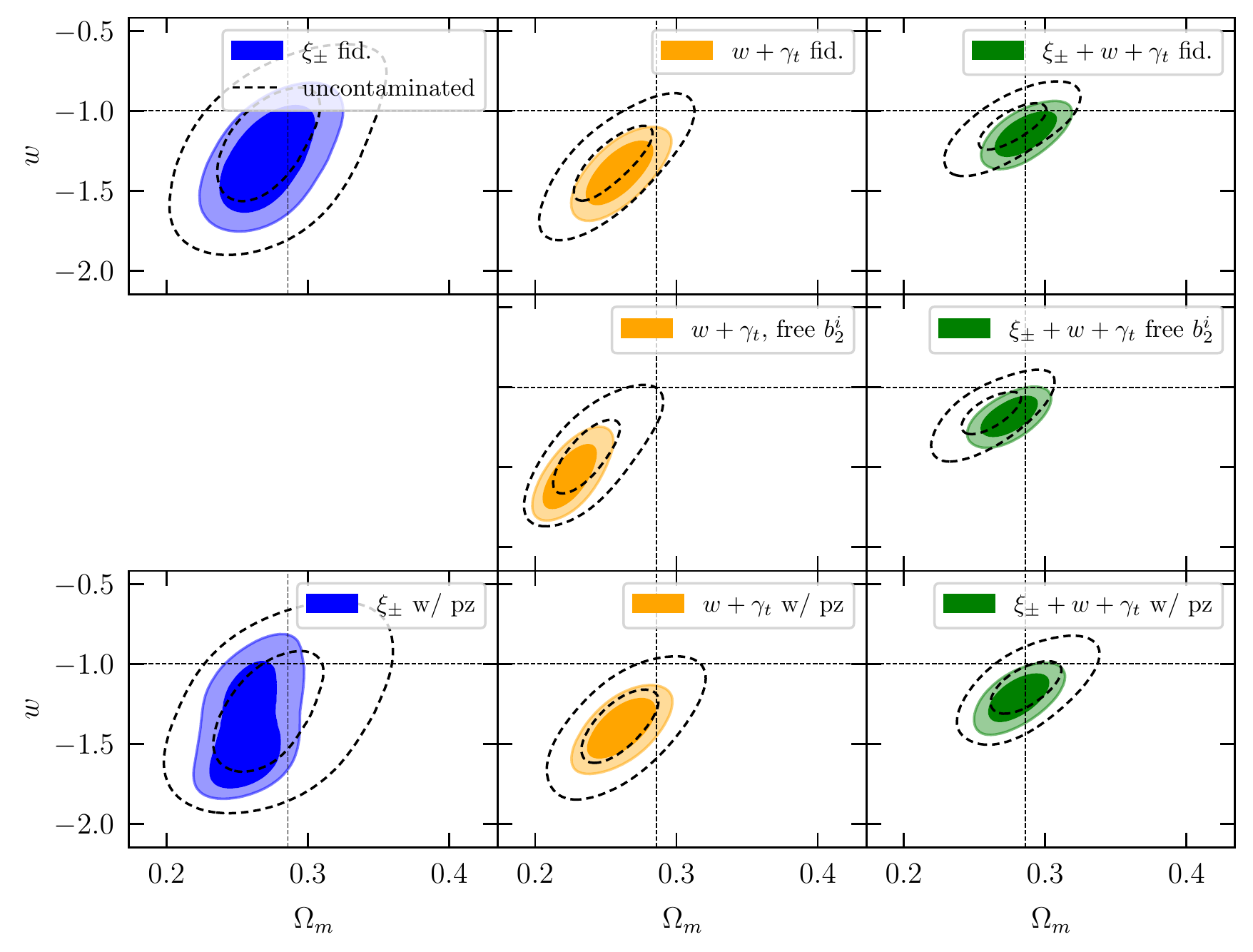}
    \caption{Same as Fig. \ref{fig:config_bcd_lcdm}, but for a $w$CDM parameter space, showing constraints on \om\ and $w$. In all cases, the probability to exceed a parameter bias of more than $0.3\sigma$ is less than $59\%$, as summarized in Tab. \ref{tab:pte}. The $w$CDM constraints in the presence of non-linear bias as shown for configuration C (middle row) are impacted by parameter projection effects, as discussed further in \cref{sec:config_c}. The inclusion of shear ratios improves the constraining power of \xipm\ alone in configuration D (bottom left) with respect to configuration B, mostly by partially breaking degeneracies with intrinsic alignment parameters, even though configuration D also marginalizes over additional nuisance parameters.}
    \label{fig:config_bcd_wcdm}
\end{figure*}

\subsection{Linear Bias}
\label{sec:config_b}
We proceed to analyze our simulations assuming the true measured redshift distributions with analysis configuration B. Configuration B samples over $\Lambda$CDM cosmological parameters, linear galaxy bias, and the TATT IA model while fixing nuisance parameters related to photometric redshift uncertainty and weak-lensing shear calibration. We also fix the parameters describing lens magnification to the values obtained for the \buzzard\ simulations in \citet*{y3-2x2ptmagnification}. 

The results of this analysis are summarized in the top row of panels in \cref{fig:config_bcd_lcdm} and \cref{fig:config_bcd_wcdm}, where we display posteriors obtained from analyses of \xipm, $\wtheta + \gammat$, and $\xipm + \wtheta + \gammat$ in the three different sub-figures for a $\Lambda$CDM and $w$CDM analysis respectively. The solid contours in these figures are obtained by running analyses on $\textbf{d}_{\rm cont}$, i.e. the mean \buzzard\ data vector, and the inner and outer contours represent the $1/\sqrt{18}\sigma$ and $2/\sqrt{18}\sigma$ confidence regions using a covariance matrix appropriate for a single simulation realization. We emphasize that these confidence regions are equivalent to the $1\sigma$ and $2\sigma$ confidence regions we would have obtained using a covariance appropriate for the mean of 18 simulations in the limit of flat priors. 

The dashed contours in each panel are constraints obtained by running on an uncontaminated data vector assuming the covariance of a single Y3 simulation, and so approximately represent the uncertainties on $S_8$, $\Omega_m$, and $w$ that we would expect from the Y3 data in the absence of photometric redshift and shear calibration systematics. The PTE values obtained by plugging these posteriors into Eq. \ref{eq:noisy_scalecut} are summarized in the first two columns of \cref{tab:pte}. All three analyses pass the criterion in Eq. \ref{eq:noisy_scalecut}. We also quote biases in the mean $\Omega_m-S_8$ and $w-\om$ posteriors for the $\Lambda$CDM and $w$CDM analyses with respect to the black uncontaminated posteriors in \cref{tab:mean_bias}. We find that all data combinations result in mean posterior biases that are less than $0.3\sigma$.

In $w$CDM, we see that the uncontaminated black-dashed posteriors are biased with respect to the truth for \twottwo\ and \ttt\ analyses. This is a so-called projection effect that results from degeneracies between our nuisance parameters and cosmological parameters, particularly $w$ and $S_8$. In configuration B, poorly constrained non-linear IA parameters in our TATT model couple with $S_8$ and $w$, such that when we marginalize over these IA parameters differences in posterior volume sourced by correlations between IA parameters and $S_8$ and $w$ lead to apparent biases. It is for this reason that it is important to quote offsets with respect to the uncontaminated posteriors, rather than the true parameter values in our simulations, lest we neglect the impact of these projection effects. The direction and size of these projection effects depends sensitively on the mean nuisance and cosmological parameter values assumed when generating our data vectors, as they are entirely a function of the parameter volume that is projected over to obtain marginalized posteriors. As such, it is also important that our data vector that is used to obtain uncontaminated posteriors is generated with nuisance parameters that match those found in the \buzzard\ simulations. 

If our forward model exactly matched the \buzzard\ measurements at the true \buzzard\ cosmology, then the projection effects in the contaminated and uncontaminated posteriors would match, as the posteriors would then agree perfectly and thus projection effects would be identical. Because we do not have an analytic forward model that exactly describes the \buzzard\ measurements, this sets a limit on the precision with which we can perform the tests presented in this work because differences in projection effects in combination with biases in our model can then source marginalized posterior offsets. This is particularly true in the limit where parameter projection effects are large, as the PTE values and parameter offsets that we quote depend on the nuisance parameter values assumed in our uncontaminated data vectors.

While these effects are counter-intuitive and and limit our ability to interpret posteriors at the precision of fractions of $1\sigma$, we can take solace in the fact that they are not a systematic bias, as the size of these effects will shrink proportionally to the constraining power of the data.

\subsection{Non-linear Bias}
\label{sec:config_c}
In configuration C we add non-linear bias to the model used in configuration B. In particular, we use the non-linear bias model described in Sec. \ref{sec:model}, and sample over $\sigma_8 b_{1}^{i}$ and $\sigma_8^2 b_{2}^{i}$ rather than the bare bias parameters in order to mitigate parameter projection effects as discussed in \cite{y3-2x2ptbiasmodelling}. We again assume the true measured redshift distributions and fix nuisance parameters related to photometric redshift uncertainty and weak lensing shear calibration. In this case, we relax our angular scale cuts so that they correspond to $r_{min}=4\, \mpc$ at the low edge of each lens redshift bin for \wtheta\ and \gammat, as motivated by \citet{Pandey2020a}. 

The results of this analysis are shown in the middle rows of \cref{fig:config_bcd_lcdm} and \cref{fig:config_bcd_wcdm} for $\Lambda$CDM and $w$CDM respectively. The posteriors here are analogous to those produced using analysis configuration B. The PTE values obtained by plugging these posteriors into Eq. \ref{eq:noisy_scalecut} are summarized in the third and fourth columns of \cref{tab:pte}. Biases in the mean $\Omega_m-S_8$ and $w-\om$ posteriors for the $\Lambda$CDM and $w$CDM analyses with respect to the black uncontaminated posteriors are quoted in \cref{tab:mean_bias}. There is a moderate decrease in $\Omega_m - S_8$ parameter bias for the full \ttt\ analysis in this configuration compared with the constraints obtained using configuration B, but the shift that we observe is approximately the same size as that expected from the tests performed in \citet{y3-2x2ptbiasmodelling}, $\sim 0.2\sigma$, that were originally used to determine the scale cuts that we are further testing here. 

In $w$CDM, we see that the PTE values and $w-\om$ parameter biases increase with respect to those found in configuration B. This is likely a result of a mismatch in projection effects occurring in our \buzzard\ analysis and our uncontaminated analysis. The first piece of evidence for this is that we see the opposite effect, i.e., a decrease in PTE and parameter biases in our configuration C $\Lambda$CDM analyses where projection effects are less important. Furthermore, we see no evidence that our non-linear bias model is a worse fit for the relaxed scale cuts used in analysis C than our linear bias model is when using our fiducial scale cuts. This is evidenced by the very small change in reduced chi-squared values between configuration A1 and A2, as described in \cref{sec:config_a}. Finally, the maximum likelihood $w$ value for the configuration C \buzzard\ \ttt\ analysis is $-1$, indicating that there are indeed large projection effects in this analysis, as this maximum likelihood value is significantly offset from the mean of the $w$ posterior in the middle right panel of \cref{fig:config_bcd_wcdm}. In light of this, we caution against over-interpretation of the larger PTE and parameter bias values for these $w$CDM constraints. This issue further emphasizes the importance of quoting parameter constraints where the effects of parameter projection effects are minimized, as has been a focus for the DES Y3 analyses.

\subsection{Calibrated photometric redshifts}
\label{sec:config_d}
Finally, we test whether our photometric redshift marginalization methodology is sufficient to recover unbiased cosmological constraints on our simulations in the presence of realistic photometric redshift uncertainties. The methodology for obtaining the calibrated photometric redshift distributions that we employ in this analysis is described in detail in \cref{sec:photoz}, as well as in \citet*{y3-sompzbuzzard,y3-sompz,Sanchez2020,y3-sourcewz,y3-lenswz,y3-shearratio}. In summary, we use lens redshift distributions as estimated by the \redmagic\ algorithm, and samples of our source redshift distributions generated by the \sdir\ algorithm \citep{Sanchez2020, y3-sompz} using the \texttt{SOMPZ} redshift distribution estimates, weighted by the likelihood of those samples given the cross-correlation of our source galaxies with \redmagic\ and spectroscopic galaxies.

We have made a significant simplifying assumption in our simulations: that we have we have sparse but unbiased spectroscopic redshift calibration samples for both \redmagic\ and our source photo-$z$ estimation methodology. This is not assumed in the Y3 analysis on data, where we have added additional uncertainty in order to encapsulate possible biases in these calibration samples.

The lens photo-$z$ uncertainties are incorporated by shifting the means of the fiducial lens $n(z)$ estimates, an approximation that has been shown to be sufficient for the \redmagic\ lens sample in \citet{y3-galaxyclustering}. We also marginalize over a re-scaling of the width of $n(z)$ for the highest redshift lens bin, as was determined necessary in \citet{y3-lenswz}. The source redshift distributions have significantly more uncertainty than our lens redshift distributions, and so we have developed methodology that enables us to explicitly marginalize over samples of this redshift distribution that encapsulate complexity that is more significant than shifts in the means of our redshift distributions. \citet{y3-hyperrank} has shown that this additional uncertainty is negligible for our DES Y3 analysis, and that marginalizing over shifts in the means of our source redshift distributions is sufficient both in the \buzzard\ simulations and in the DES Y3 data. As such, for this analysis we marginalize over a Gaussian prior on these mean shift parameters. Priors on all photometric redshift related parameters are listed in \cref{tab:params}. In particular, we assume values for these priors based on preliminary characterizations of these effects in the DES Y3 data, although we note that the size of these priors contributes negligibly to the size of our posteriors, as IA- and bias-related nuisance parameters are our dominant systematics.

Additional information on our redshift distributions as well as IA parameters is gained through the inclusion of small-scale shear-ratio measurements at the likelihood level, as described in \cref{sec:shearratio} and \citet{y3-shearratio}. We also marginalize over multiplicative shear biases in this analysis configuration, with uncertainties determined by a preliminary estimate of these on our Y3 data \citep{y3-imagesims}. These multiplicative biases have no discernible impact on our posteriors due to their very small uncertainties. 

Results of this analysis are summarized in the bottom rows of \cref{fig:config_bcd_lcdm,fig:config_bcd_wcdm}, where now we compare the posteriors obtained using analysis configuration D (solid) with those obtained with configuration B (dashed). The dashed contours are now approximately the $0.3\sigma$ and $1\sigma$ constraining power of a single Y3 analysis. We choose to compare configuration D to configuration B rather than an uncontaminated analysis because we wish to isolate parameter biases that are sourced by photometric redshift (mis)estimation. 
We see that in all cases the shifts in posteriors are negligible, indicating that our photometric redshift calibration methodology has been successful in encapsulating the photo-$z$ biases contained in our simulations. PTE values and parameter biases for analysis configuration D are listed in \cref{tab:pte} and \cref{tab:mean_bias}. 

\section{Summary and Conclusions}
\label{sec:conclusion}
We have presented the \buzzardtwo\ simulations, a suite of 18 synthetic DES Y3 galaxy catalogs tailored for the validation of combined clustering and lensing analyses. We have used these simulations to test the robustness of a number of choices made in the DES Y3 \ttt\ analysis, showing in particular that:
\begin{enumerate}
    \item Our model can fit the measurements from our simulations at the simulations' true cosmology.
    \item Using the true redshift distributions in our simulations and sampling over cosmology, linear bias, and intrinsic alignments, we can recover the true cosmology of our simulations using cosmic shear alone, \twottwo\ and \ttt\ analyses.
    \item Using the true redshift distributions in our simulations and sampling over cosmology, \textit{non-linear} bias, and intrinsic alignments, we can recover the true cosmology of our simulations, using \twottwo\ and \ttt\ analyses.
    \item We are able to recover unbiased cosmological constraints when assuming calibrated photometric redshift distributions, making use of the full calibration methodology applied to the Y3 data.
\end{enumerate}

In Sec. \ref{sec:buzzsum} we describe the new suite of \buzzard\ simulations used in this work and elsewhere in our DES Y3 analyses, highlighting the improvements that have been made over the simulations used in DES Y1. In particular, we demonstrate improved agreement between color and magnitude distributions for our source and lens galaxy samples in \cref{fig:magdist,fig:colors,fig:red_sequence}. 

In Sec. \ref{sec:datavec} we describe how we measure our \ttt\ data vectors, including the photometric redshift calibration methodology that was applied to the simulations. We present comparisons between redshift distributions for source and lens galaxies (\cref{fig:nofz}) and the \ttt\ measurements (\cref{fig: 2x2_data_sims,fig: xipm_data_sims}), showing significant improvements in the level of agreement from similar comparisons in DES Y1. We also highlight the four-step calibration that is applied to the source photometric redshift distributions in our simulations: SOM based photometric redshift estimation via \texttt{SOMPZ} and \sdir, clustering redshifts, and shear ratios. We have kept as close to the procedure employed in the DES Y3 data as possible, and we note the important aspects of these algorithms that were validated using the \buzzard\ simulations that are presented in greater detail in \citet*{y3-sompz,Alarcon2020,Sanchez2020,y3-lenswz,y3-sourcewz,y3-shearratio}.

In Sec. \ref{sec:model} we describe the models applied to the simulations, and in Sec. \ref{sec:validation} we show that these models produce constraints on $S_8$, $\Omega_m$ and $w$ that are biased at $<0.3\sigma$ ($1\sigma$) with a probability of at least $38\%$($98\%$) in $\Lambda$CDM and $42\%$($99\%$) in $w$CDM, while accounting for residual noise in the measurements from our simulations. These results are summarized in \cref{tab:pte} and \cref{fig:config_bcd_lcdm,fig:config_bcd_wcdm}. Mean two-dimensional parameter biases are less than $0.3\sigma$ for all analysis configurations and are summarized in \cref{tab:mean_bias}. These results demonstrate that our \ttt\ analysis is robust to the assumptions made regarding bias modeling, non-linearities in the matter distribution, higher-order lensing effects, non-Gaussianity of the likelihood function, and approximations made in photometric redshift estimation in a realistic simulated analysis setting.

We note that the probabilities quoted above would asymptote to either zero or one in the limit of infinite simulations, which suggests the question: why we have not made an effort to generate more simulations in order to more precisely determine our systematic biases? The reason for this choice is partially pragmatic, as generating these simulations requires significant computational and human time. An equally important reason is that the analyses presented in this work suffer from systematic errors that contribute non-negligible uncertainty to these probabilities. 

One such systematic, namely interpretation of posterior probability distributions in the presence of projection effects, has already been discussed and argued to be important for many of the analyses presented here. Significant effort has been made to minimize projection effects in these analyses. For example, where ever possible we have reduced the complexity of our models by removing additional parameters that are not required at the precision of our data. Although these projection effects limit the interpretability of our posteriors at the fraction of $1\sigma$ level, we reiterate that they are not a systematic bias in the traditional sense, as the size of these effects will decrease proportionally to the constraining power of our analyses.

In addition to systematics intrinsic to our posterior distributions, systematics in our simulations are also important. It is necessary to make simplifying assumptions when generating large suites of simulations as we have done here, and ruling out the contribution of these assumptions to the parameter offsets in this analysis is a time intensive task. Efforts to systematize these tasks \citep{Mao2017} are extremely important to the continued ability to perform analysis validation tests such as those presented in this work.

A number of important effects have been left out of these simulations, including shear calibration biases, spectroscopic incompleteness in photo-$z$ calibration, more realistic survey inhomogeneity, galaxy intrinsic alignments, and baryonic effects on the matter distribution. The treatment of these effects is thoroughly validated elsewhere for our DES Y3 \ttt\ analysis, but work is ongoing to incorporate many of these into future versions of these simulations. With significant investment, we anticipate that improvements in the methodologies used to generate these simulations will continue to meet the validation needs of upcoming joint clustering and lensing analyses.

\begin{acknowledgments}
JD is supported by the Chamberlain fellowship at Lawrence Berkeley National Laboratory. This research used resources of the National Energy Research Scientific Computing Center (NERSC), a U.S. Department of Energy Office of Science User Facility located at Lawrence Berkeley National Laboratory, operated under Contract No. DE-AC02-05CH11231. Some of the computing for this project was performed on the Sherlock cluster. We would like to thank Stanford University and the Stanford Research Computing Center for providing computational resources and support that contributed to these research results.

Funding for the DES Projects has been provided by the U.S. Department of Energy, the U.S. National Science Foundation, the Ministry of Science and Education of Spain, 
the Science and Technology Facilities Council of the United Kingdom, the Higher Education Funding Council for England, the National Center for Supercomputing 
Applications at the University of Illinois at Urbana-Champaign, the Kavli Institute of Cosmological Physics at the University of Chicago, 
the Center for Cosmology and Astro-Particle Physics at the Ohio State University,
the Mitchell Institute for Fundamental Physics and Astronomy at Texas A\&M University, Financiadora de Estudos e Projetos, 
Funda{\c c}{\~a}o Carlos Chagas Filho de Amparo {\`a} Pesquisa do Estado do Rio de Janeiro, Conselho Nacional de Desenvolvimento Cient{\'i}fico e Tecnol{\'o}gico and 
the Minist{\'e}rio da Ci{\^e}ncia, Tecnologia e Inova{\c c}{\~a}o, the Deutsche Forschungsgemeinschaft and the Collaborating Institutions in the Dark Energy Survey. 

The Collaborating Institutions are Argonne National Laboratory, the University of California at Santa Cruz, the University of Cambridge, Centro de Investigaciones Energ{\'e}ticas, 
Medioambientales y Tecnol{\'o}gicas-Madrid, the University of Chicago, University College London, the DES-Brazil Consortium, the University of Edinburgh, 
the Eidgen{\"o}ssische Technische Hochschule (ETH) Z{\"u}rich, 
Fermi National Accelerator Laboratory, the University of Illinois at Urbana-Champaign, the Institut de Ci{\`e}ncies de l'Espai (IEEC/CSIC), 
the Institut de F{\'i}sica d'Altes Energies, Lawrence Berkeley National Laboratory, the Ludwig-Maximilians Universit{\"a}t M{\"u}nchen and the associated Excellence Cluster Universe, 
the University of Michigan, NFS's NOIRLab, the University of Nottingham, The Ohio State University, the University of Pennsylvania, the University of Portsmouth, 
SLAC National Accelerator Laboratory, Stanford University, the University of Sussex, Texas A\&M University, and the OzDES Membership Consortium.

Based in part on observations at Cerro Tololo Inter-American Observatory at NSF's NOIRLab (NOIRLab Prop. ID 2012B-0001; PI: J. Frieman), which is managed by the Association of Universities for Research in Astronomy (AURA) under a cooperative agreement with the National Science Foundation.

The DES data management system is supported by the National Science Foundation under Grant Numbers AST-1138766 and AST-1536171.
The DES participants from Spanish institutions are partially supported by MICINN under grants ESP2017-89838, PGC2018-094773, PGC2018-102021, SEV-2016-0588, SEV-2016-0597, and MDM-2015-0509, some of which include ERDF funds from the European Union. IFAE is partially funded by the CERCA program of the Generalitat de Catalunya.
Research leading to these results has received funding from the European Research
Council under the European Union's Seventh Framework Program (FP7/2007-2013) including ERC grant agreements 240672, 291329, and 306478.
We  acknowledge support from the Brazilian Instituto Nacional de Ci\^encia
e Tecnologia (INCT) do e-Universo (CNPq grant 465376/2014-2).

This manuscript has been authored by Fermi Research Alliance, LLC under Contract No. DE-AC02-07CH11359 with the U.S. Department of Energy, Office of Science, Office of High Energy Physics.
\end{acknowledgments}

\appendix

\section{Color-dependent clustering}
\label{app:colordepclustering}
We impart a color-dependent clustering signal to our simulations in a two-step manner. First, SEDs are assigned to each simulated galaxy by finding a galaxy in the SDSS Main Galaxy Sample (SDSS MGS) with a close match in $M_r$, and assigning its SED to the simulated galaxy, preferentially choosing blue SEDs over red ones proportional to $\frac{f_{\rm red}(z)}{f_{\rm red}(z=0)}$, where $f_{\rm red}(z)$ is the fraction of red galaxies found at redshift $z$ as described in Appendix E.2 in \citet{DeRose2019}. Each SED is represented in our simulations by a set of 5 \kcorrect\ templates \citep{Blanton2005}. 

Once each galaxy has an SED, we perform a conditional abundance matching procedure. In particular, for every galaxy we compute $R_{h}$, the distance to the nearest halo above a mass cut of $M_{h,{\rm cut}}$. We then shuffle SEDs between galaxies in order to enforce the relation

\begin{align}
\label{eq:cam}
    P(< g-r | M_r) = P(<R_{h} | M_r),
\end{align}
where $g-r$ is the rest frame $g-r$ color of each galaxy. In practice, we introduce an extra parameter in this model to allow for a non-unity correlation, $r$ between $R_h$ and $g-r$, as described in Eqs. 12-13 in \citet{DeRose2021}. We use the best-fit values of $r$ and $M_{h,{\rm cut}}$ from \citet{DeRose2021}, where this model is fit to SDSS MGS redshift space clustering measurements. This procedure makes use of conditional abundance matching algorithms implemented in \citet{Hearin2016}.

\section{Red-sequence color model}
\label{app:redsequence}
We match the mean and scatter of the red-sequence in our simulations to that observed in DES Y3 data by applying the following algorithm:

\begin{algorithm}[H]
  \begin{algorithmic}
    \For{galaxy $g$ with redshift $z$ and absolute magnitude $M_{r}$}
    \If{$g-r > 0.095 - 0.035\,M_{r}$}
    \For{band $b\: \in\: \{r,i,z\}$}
        \State Add mean offset, $\langle \Delta c_{b} (z)\rangle$, to magnitude $m_{b}$
        \State Add noise with variance $\Delta \sigma_{b}(z)^2$ to $m_b$
    \EndFor
    \EndIf
    \EndFor
  \end{algorithmic}
\end{algorithm}

Here, $\langle \Delta c_{b}(z)\rangle$ and $\Delta \sigma_{b}^2$ are the mean offset in in red-sequence color and difference between scatter in the red-sequence as a function of $z$ between the DES Y3 data, and the unaltered version of these simulations. Since there are four photometric bands in DES, but only three unique colors, we can assign the mean differences in $g-r$, $r-i$, and $i-z$ to $r$, $i$, and $z$ without loss of generality. When adding additional noise to the magnitudes in our simulations, we must account for the fact that noise in a band $b$ contributes to two colors $X-b$ and $b-Y$, where $X$ and $Y$ are adjacent bands to $b$. Thus, in order to match the width of $P(X-b | z)$ and $P(b-Y | z)$ we must take $\Delta \sigma_{b}(z)^2 = \sigma(X-b | z)^2 - \sigma(b-Y | z)^2$. This procedure leads to very good matches to the mean and scatter of the red sequence observed in DES Y3 as shown in Fig. \ref{fig:red_sequence}. 

\section{Photometric error model}
\label{app:errormodel}
We make use of the relationship between true photometry and noisy wide-field photometry as measured by \balrog\ to add photometric noise to our simulations. The algorithm for doing so is as follows:

\begin{algorithm}[H]
  \begin{algorithmic}[1]
    \For{each simulated galaxy $g$}
    \State Find galaxy $g^{\prime}$ in DES Y3 deep fields by matching $\mathbf{m_{g}}$
    \State Randomly choose a wide-field injection $\hat{g^{\prime}}$ of $g^{\prime}$
    \If{$\hat{g^{\prime}}$ was not detected }
    \State Set all of $g$'s observed magnitudes to 99
    \Else 
    \State Compute error in the wide-field magnitudes $\Delta \mathbf{m_{g^{\prime}}}$
    \State Add $\Delta \mathbf{m_{g^{\prime}}}$ to true simulated magnitudes $\mathbf{m_{g}}$
    \EndIf
    \EndFor
  \end{algorithmic}
\end{algorithm}

\noindent where $\mathbf{m_{g}}$ are $riz$ magnitudes of the simulated galaxy $g$, and $\Delta \mathbf{m_{g^{\prime}}}$ is the difference between the true injected $riz$ magnitudes from $g^{\prime}$ and the magnitudes measured from the wide-field injection $\hat{g^{\prime}}$. In this way, we reproduce the relation $p(\Delta \mathbf{m_{g}} | \mathbf{m_{g}})$ measured in the DES Y3 data exactly. 

\section{DES Y3 Source and Lens Galaxy Sample Selection}
\label{sec:sample_selection}
In this work we focus on reproducing the properties of the \metacal source galaxy sample \citep{y3-shapecatalog} and \redmagic\ lens galaxy sample \citep{y3-galaxyclustering} in order to mimic the fiducial DES Y3 \ttt\ analysis. The \redmagic\ galaxy sample is selected using wide-field $griz$ photometry, and so we are able to apply the same selection algorithm described in \citet{Rozo2015} and \citet{y3-galaxyclustering} to our simulated galaxy catalogs. In particular, we select two different \redmagic\ catalogs, that we refer to as the \texttt{HighDens} and \texttt{HighLum} samples. 

\begin{figure}
	\includegraphics[width=\columnwidth]{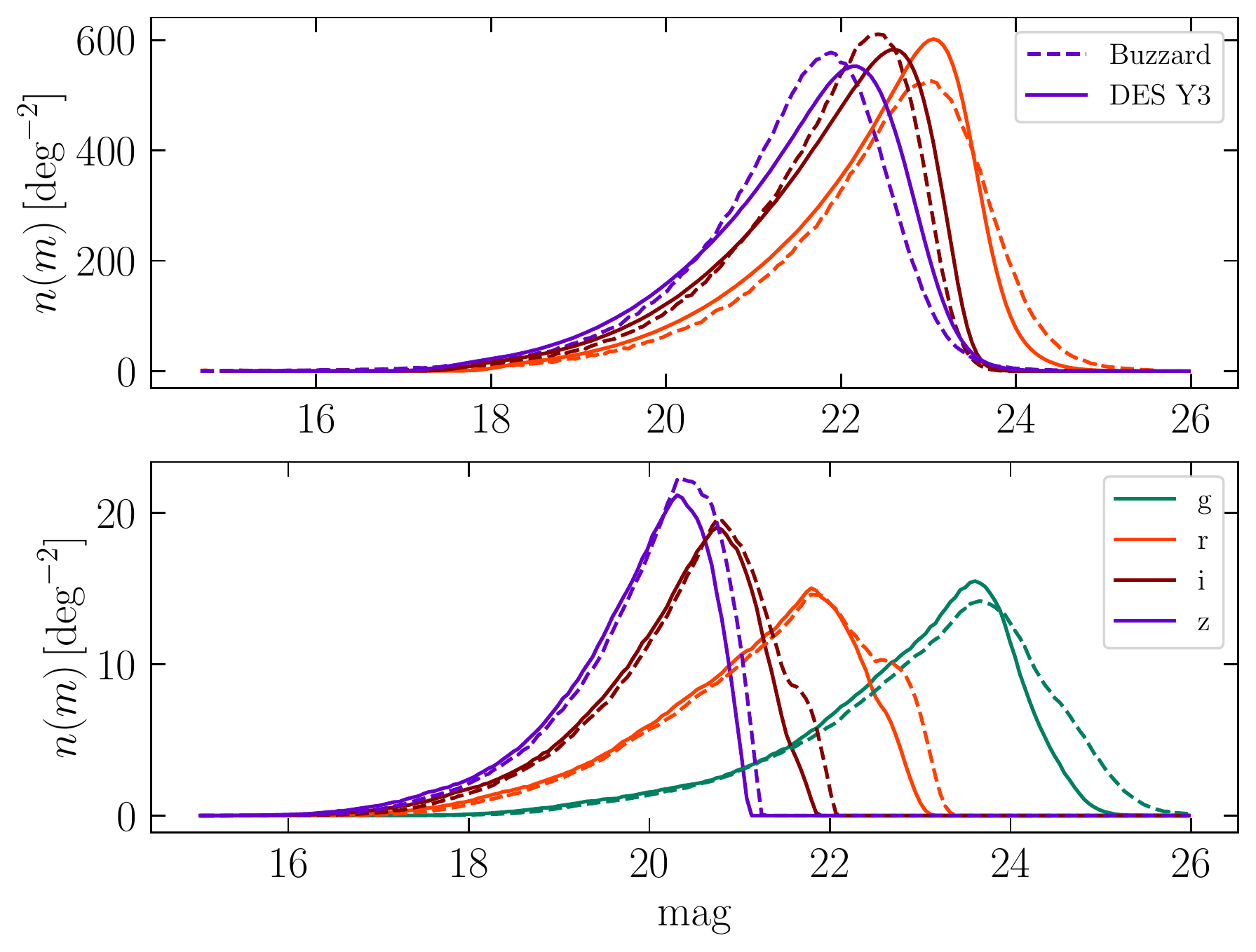}
    \caption{Comparison of magnitude distributions for sources (left) and \redmagic\ lenses (right) between Buzzard (dashed) and the DES Y3 data (solid). 
    Different colors represent different photometric bands, where we compare $riz$ for sources and $griz$ for lenses, because these are the bands that are used for each respective selection. Overall agreement is good, with fractional differences between \buzzard\ and the DES Y3 data not exceeding $20\%$.}
    \label{fig:magdist}
\end{figure}

The first of these has the highest number density, and is used for the first three lens bins of our \ttt\ data vector. The second is lower number density but can be selected out to $z<0.9$, and so is used for our two higher redshift lens bins. Applying this procedure results in magnitude and photometric redshift distributions that are nearly identical to those found in the DES Y3 data, as shown in the right panels of \cref{fig:magdist,fig:nofz}. 

Selecting a source galaxy sample that matches the properties of the DES Y3 \metacal sample is more challenging. This is partially due to the fact that this sample pushes up against the detection limit of the wide-field survey and so includes implicit selection on many properties such as surface brightness, morphology, proximity to bright objects, and observing conditions that are not modeled in our simulations. This is also a fainter sample than \redmagic\ in general and so involves galaxies in our simulations that are more sparsely observed in the spectroscopic samples that we use to inform our simulated galaxy samples. Given these limitations, we focus on constructing a simulated source sample that matches the effective number density, shape noise, and redshift distributions of the DES Y3 \metacal sample, as these are the properties that largely govern the cosmological constraining power of the cosmic shear and galaxy--galaxy lensing measurements that we wish to analyse.

In order to do this we apply two cuts to our simulated galaxy sample motivated by cuts that are performed to construct the DES Y3 \metacal sample:

\begin{enumerate}
    \item $10 < f_{i}/\sigma(f_{i}) < 1000$, and 
    \item $r_{\rm eff} / r_{\rm psf} > x_1 / (1 + x_2 z) + x_3$.
\end{enumerate} 

One of the main cuts that influences the number density of the Y3 \metacal sample is a cut on signal-to-noise ratio (SNR), $10 < \textrm{SNR} < 1000$. The first cut above approximates this SNR cut, where we compute \textrm{SNR} as $\textrm{SNR} = f_{i}/\sigma(f_{i})$ and $f_{i}$ is the noisy simulated $i$-band flux and $\sigma(f_{i})$ is the error in that flux, computed as described in App. \ref{app:errormodel}. Applying only this cut results in a sample with an over-abundance of galaxies, so we apply an additional cut on galaxy size, motivated by the cut in the ratio of galaxy size to PSF size applied to the Y3 data. Here we approximate the Gaussian size used in the Y3 data with the simulated half-light radius $r_{\rm eff}$ where $r_{\rm eff}=\sqrt{r_{\rm 50}^2 + r_{\rm psf}^2}$ is the PSF convolved galaxy half-light radius, $r_{\rm 50}$, and the PSF size, $r_{\rm psf}$, is taken from the DES Y3 maps of PSF size at the position of the simulated galaxy. The $x_{i}$ are free parameters in this cut that we fit to the total response and inverse variance weighted number density of a preliminary version of the DES $Y3$ \metacal catalog, $n_{\rm eff}=5.9$. This fit yields a number density in our simulations of $n_{\rm eff}=5.84$. Before measuring correlation functions with this source catalog, we also explicitly match the effective shape noise, $\sigma_{e}=[0.247, 0.266, 0.263, 0.314] $, in the four tomographic bins used. This procedure leads to good matches to the magnitude and redshift distributions observed in the DES Y3 \metacal sample as shown in \cref{fig:magdist,fig:nofz}. The photometric redshift estimation methodology for the simulated source sample is discussed at more length in Sec. \ref{sec:photoz} and \citet{y3-sompz}.

\section{Two-point function estimation}
\label{sec:twopoint_est}
To compute the galaxy angular auto-correlation function for a single tomographic bin, $w(\theta)$, we use the Landy--Szalay estimator \cite{LandySzalay}
\begin{align*}
    \hat{w}(\theta) &= \frac{DD - 2DR + RR}{RR} \\
    &= \sum_{ab} \frac{w_{a}w_{b}}{N_{L}(N_{L} - 1)} - 2\sum_{aR} \frac{w_{a}}{N_{L}N_{R}} \\
    & + \sum_{RR} \frac{1}{N_{R}(N_{R} - 1)} .
\end{align*}\\ 
The first sum runs over pairs of lens galaxies separated by an angle $\theta_{\rm min} < \theta < \theta_{\rm max}$, where $\theta_{\rm min}$ and $\theta_{\rm max}$ are the edges of the angular bin, $w_{a/b}$ are the systematic weights associated with galaxies $a$ and $b$ and $N_L$ is the total number of lens galaxies in the tomographic bin under consideration. The lens weights are determined using the same algorithm described in \cite{y3-galaxyclustering}. The second and third sums run over lens--random and random--random pairs in the same angular bin, and $N_R$ is the total number of randoms.  We always take this to be $N_R=20N_L$, as opposed to the $N_R=40N_L$ used in the measurements on the data. We work with the mean measurement from many simulations, so extra noise resulting from the use of fewer randoms per lens galaxy is offset by this fact and allows us to save a significant amount of computing time when measuring angular clustering in the \buzzard\ simulations.

Following the choices made in the DES Y3 data, we bin the \redmagic\ sample into five distinct redshift bins, with edges $z=\{0.15, 0.35, 0.5, 0.65, 0.8, 0.9\}$. Galaxies are divided between these bins using the mean of the \redmagic\ photometric redshift PDF, $z_{\rm mean, \redmagic\ }$. For the first three bins, we use the \redmagic\ \texttt{HighDens} sample; the two highest redshift bins use the \redmagic\ \texttt{HighLum} sample. 

Our estimator for tangential shear around lens galaxies includes boost factors and random point subtraction as follows:
\begin{align}
\hat{\gamma}_t (\theta) &= \hat{\gamma}_{t,\, \rm no\, boosts} (\theta)B(\theta) - \hat{\gamma}_{t,\, \rm rand} \label{eq:gammat_term_est}\\
&= \frac{N_r}{N_l} \frac{\sum_{LS} w_{L} \, \epsilon_{t, LS}(\theta)}{N_{\rm rs}} - \frac{\sum_{RS}\epsilon_{t, RS}(\theta)}{N_{\rm rs}}\, , 
\label{eq:gammat_est}
\end{align}
where in Eq. \ref{eq:gammat_term_est} we have written the estimator in terms of the uncorrected tangential shear estimator $\hat{\gamma}_{t,\, \rm no\, boosts}$, the boost factor $B(\theta)$, and the estimator for tangential shear around random points $\hat{\gamma}_{t,\, \rm rand}$. We further expand this in Eq. \ref{eq:gammat_est}, where $N_r$ is the number of randoms, $N_l = \sum w_{L}$ is the effective number of lenses given by the sum over the lens weights for all lens galaxies $w_l$, $\epsilon_{t, LS}$ is the tangential shear of a source-lens pair, $\epsilon_{t, RS}$ is the tangential shear of a source-random pair, and $N_{\rm rs}$ is the number of random-source pairs. This expression is a simplified version of that shown in \citet{y3-gglensing}, where we have set all \metacal responses, as well as source and random weights to 1. We measure $\gamma_t$ for each of four source galaxy tomographic bins, as described in Sec. \ref{sec:photoz}, around the five tomographic lens galaxy bins, resulting in a total of 20 unique cross-correlation measurements.

The estimator of the shear--shear correlation functions can be written in terms of the measured radial, $\epsilon_{\rm x}$, and tangential, $\epsilon_{\rm t}$, components of ellipticities defined per galaxy pair along the line of separation between the galaxies: 
\begin{equation}
\hat{\xi}_{\pm}(\theta)= \langle\epsilon_t\epsilon_t \pm \epsilon_{\times}\epsilon_{\times} \rangle (\theta) \,.
\end{equation}
This is determined by averaging over all galaxy pairs $(a, b)$ separated by an angle $\theta$ as
\begin{equation}
\label{eqn:2ptcorrfnestimator}
\centering
\hat{\xi}_{\pm}(\theta) = \frac{\sum_{ab}[\epsilon_{\rm t, a} \epsilon_{\rm t, b} \pm \epsilon_{\times, a}\epsilon_{\times, b}]}{N_{\rm pair}} \, , 
\end{equation}
where $N_{\rm pair}$ is the number of galaxy pairs separated by an angle $\theta$. This is again a simplified version of the expression used in the measurements on the DES Y3 data, as we do not use \metacal responses or inverse variance weights in our simulations. We measure all 10 unique auto- and cross-correlations of the four tomographic source galaxy bins.

In all cases, correlation functions are measured in 20 logarithmically spaced bins between 2.5 and 250 arcmin, and the mean angle of the bin is reported as the pair-count averaged separation within that bin. We describe our redshift binning algorithms for our source and lens galaxy samples in the following sub-section.

\section{Photometric Redshift Calibration}
\label{sec:photoz}

Compared with our source galaxy photometric redshift estimation algorithm, our lens galaxy photometric redshift estimation is relatively unchanged from that used in DES Y1. We briefly describe it here, and refer the reader to \citet{Rozo2015}, \citet{y3-lenswz} and \citet{y3-galaxyclustering} for more details. As the \redmagic\ sample is a set of bright galaxies with abundant spectroscopy, we place significant confidence in the \redmagic\ photometric redshift estimates provided by the algorithm itself, $p(z_{\redmagic})$. These are obtained by constructing a red-sequence spectral template from a combination of spectroscopy and galaxy cluster members. In our simulations, we assume that we have a sparse but unbiased spectroscopic training set of similar size to that used in the data. We bin lens galaxies into five tomographic bins with edges, $\{0.15, 0.35, 0.5, 0.65, 0.8, 0.9\}$, using the mean of $p(z_{\redmagic})$. To estimate the $n(z)$ for each tomographic bin, we stack four Monte-Carlo samples drawn from the $p(z_{\redmagic})$ for each galaxy. 

We now describe how the source galaxy photometric redshift calibration is performed in our simulations, highlighting important similarities and differences to what is done in the data. As this methodology is new to DES Y3, much of it was developed and tested against the simulations presented here. The calibration methodology, excluding calibration internal to the \ttt, is broken into four distinct steps described in the following four sections. These individual steps are described in significantly greater detail in \citet{y3-sompzbuzzard}, \citet{y3-sompz}, \citet{Alarcon2020}, \citet{Sanchez2020}, \citet{y3-sourcewz} and \citet{y3-shearratio}.

\subsection{SOMPZ}
We aim to determine the redshift distribution $n(z)$ of the source galaxy sample, proportional to the probability, $p(z)$, that a galaxy in that sample is at redshift $z$. We estimate $n(z)$ by re-weighting the redshift distribution of a sample with known redshift information in a suitable way. In the DES Y3 data, this re-weighting is performed in two steps using three different galaxy samples: a sample of wide-field galaxies that form our weak-lensing source sample (the \textit{weak-lensing source} sample), a set of galaxies with deep, many-band photometry (the \textit{deep-field} sample), and a sample with deep, many-band photometry as well as securely determined redshifts (the \textit{redshift} sample).

First, we characterize the redshift distributions of low noise galaxy detections from our deep-field photometry where we have additional photometric information in the form of near-infrared photometric measurements in $yJHK$ from the Ultra Vista survey. We discretize the $ugrizYJHK$ color space spanned by this deep-field sample using a self-organizing map (SOM) \citep{Kohonen1982}. We then estimate the redshift distribution, $p(z | c)$ in each cell, $c$, of the so-called deep SOM by stacking redshift estimates from our redshift sample. This sample is a combination of spectroscopic surveys and COSMOS+PAU galaxies with photometric redshifts \citep{Alarcon2020, Laigle2016}. In our simulations we construct analogs of these deep-field photometric catalogs by selecting patches of a single \buzzard\ simulation with the same area as the deep-field catalogs in the DES Y3 data. We apply a constant level of photometric noise to these catalogs, derived from the median depth in each deep field and construct the deep SOM from this photometry. In order to estimate $p(z | c)$ we assume that we have a redshift sample that is free of selection biases and is the same size as the redshift sample used to estimate $p(z | c)$ in the DES Y3 data.

Analogously, we construct a wide-field SOM using our entire wide-field galaxy catalog, labeling cells in this SOM as $\hat{c}$. We wish to estimate $p(z | \vec{m})$ where $\vec{m}$ is a vector of wide-field magnitudes, or in the same SOM language, $p(z | \hat{c})$, where $\hat{c}$ is the SOM cell, or phenotype, that $\vec{m}$ is placed in. This process is complicated by the fact that the wide-field sample has significantly larger photometric uncertainties than the deep-field sample, and is not supplemented by the near-IR photometry. As such, we must connect the magnitudes that we measure for our wide-field sample to those measured in our deep fields. In the data this is performed by injecting deep-field galaxies into wide-field observations with \balrog\ allowing for the estimation of $p(c | \hat{c})$, i.e., the probability that a wide-field galaxy that is placed in the wide-field SOM cell $\hat{c}$ would be placed in the deep-field SOM cell $c$, had it been observed in the deep fields rather than the wide field. In our simulations, we can estimate $p(c | \hat{c})$, by determining which deep SOM cell $c$ each wide-field galaxy in cell $\hat{c}$ falls into. This relation is determined using the same number of galaxies in our simulations as we have \balrog\ injections in DES Y3 data.

Once galaxies have been assigned to cells $\hat{c}$ based on their photometric information, we construct tomographic bins and assign each cell to a bin. For our fiducial redshift distributions, we construct these bins according to the following procedure:

\begin{enumerate}
\item{To construct a set of $n$ tomographic bins $\hat{b}$, begin with an arbitrary set of $n+1$ bin edge values $e_j$.}
    
\item{Assign each galaxy in the redshift sample to the tomographic bin $\hat{b}$ that contains the median of its $p(z)$ (or its spectroscopic redshift $z$, if it has one). This yields a set of $N_{\mathrm{spec}, (\hat{c},\hat{b})}$ galaxies satisfying the dual condition of membership in a wide SOM cell $\hat{c}$ and a tomographic bin $\hat{b}$. This can be written as a sum over \balrog\ realisations $i$ of redshift galaxies:}
    
\begin{equation}
    N_{\mathrm{spec}, (\hat{c},\hat{b})} = \sum_{i} \delta_{\hat{c}, \hat{c}_i} \delta_{\hat{b}, \hat{b}_i}
\end{equation}
    
\item{Assign each wide cell $\hat{c}$ to the bin $\hat{b}$ to which a majority of its constituent redshift sample galaxies are assigned:}
    
\begin{equation}
    \hat{b} = \{ \hat{c} | \mathrm{argmax}_{\hat{b}} N_{\mathrm{spec}, (\hat{c},\hat{b})}\}
\end{equation}
    
\item{Adjust the edge values $e_j$ such that the numbers of galaxies in each tomographic bin $\hat{b}$ are approximately equal and repeat the procedure from step (ii) with the final edges $e_j$.}
\end{enumerate}

After completing this procedure, our final bin edges are $z=$[0.0, 0.358, 0.631, 0.872, 2.0] for the Y3 weak lensing source catalog. Due to slight differences in the Y3 source galaxy catalogue and the simulated Buzzard equivalent, the bin edges in the equivalent Buzzard catalogue are $z=$[0.0, 0.346, 0.628, 0.832, 2.0].

Finally, we combine all these pieces of information together into the redshift distribution for each bin $\hat{b}$:

\begin{align} 
p(z|\hat{b}, \hat{s}) &\approx \sum_{\hat{c} \in \hat{b}} \sum_{c} p(z|c, \hat{b}, \hat{s}) p(c|\hat{c},\hat{s}) p(\hat{c}|\hat{s})\label{eqn:bincond}
\end{align}

\subsection{3sDir}

Once we have the SOMPZ formalism in place, we can estimate the redshift distribution of a tomographic bin using Equation~\ref{eqn:bincond}. An alternative way of writing that equation, highlighting which sample is used to inform each term, is
\begin{equation} \label{eqn:redshift_prob_samples}
    p(z|\hat{b}, \hat{s}) \approx \sum_{\hat{c} \in \hat{b}} \sum_{c} \underbrace{p(z|c)}_\text{Redshift} \underbrace{p(c)}_\text{Deep} \underbrace{\frac{p(c,\hat{c})}{p(c)p(\hat{c})}}_\text{Balrog}  \underbrace{p(\hat{c})}_\text{Wide}, 
\end{equation}
where the $\hat{b}, \hat{s}$ selections are implicit in the RHS terms. One of the main uncertainties on our estimate of $n(z)$ comes from the sample variance and shot noise present in the deep and redshift samples, which inform the probabilities $p(z|c)$ and $p(c)$. We denote the joint probability of redshift and color informed by the deep and redshift galaxy samples as a set of coefficients $\{f_{zc}\}$, with $0\leq f_{zc} \leq 1$ and $\sum_{zc} f_{zc} = 1$, where $z$ represents a redshift bin, and $c$ a deep SOM color cell.  

We implement the \sdir method, an approximate model that produces samples of $\{f_{zc}\}$ given the observed number counts of redshift and color from the deep and redshift samples including the uncertainty from shot noise and sample variance \citep{y3-sompz}. This method was developed and validated first in simulations \citep{Sanchez2020}, but for a non-tomographic sample with a different selection than the DES Y3 source sample, where all galaxies in the deep fields had redshift information, and without a transfer function to re-weight the colors in the deep field. Here, we test this method with simulations tailored to the DES Y3 samples, using an extended version of \sdir described in \citet{y3-sompz}. For each sample of the coefficients $\{f_{zc}\}$ we can compute a sample of the redshift distribution in each tomographic bin, propagating the uncertainty to the full shape of the $n(z)$. 

To test the performance of the \sdir method we perform a similar test to what was used to validate the SOMPZ algorithm, using the 300 \buzzard\ realizations of the deep fields. In each realization, we draw $10^4$ samples of the coefficients $\{f^{i}_{zc};\, i=1,\ldots,10^4\}$ using \sdir and the measured number counts of redshift and color in this realization. From it, we estimate the mean redshift of each $f^{i}_z$ sample, $\zbar^{i}=\sum_z z f^{i}_z$, and its average value $\zbar^{\sdir}\equiv \langle \zbar^{i}\rangle$ in each \buzzard\ realization. We also compute the $\zbar^{\text{SOMPZ}}$ value of the single $n(z)$ from SOMPZ in each realization, which we obtain by fixing the probabilities to the number counts. We are able to verify that the expected value of the mean redshift across the 300 realizations agrees between \sdir and SOMPZ. We also find the pull distribution between individual $\zbar$ samples from \sdir and the fiducial SOMPZ $\zbar$ to be very close to a Gaussian with zero mean and unit variance.

Note that we are changing how we parameterize the uncertainty: instead of fixing the $n(z)$ and modeling the uncertainty with a shift to the distribution, we are modeling the uncertainty in $p(z,c)$ observed in the deep fields, and fully propagating it to an uncertainty on the shape of the $n(z)$. Therefore, it is particularly important to show that no significant biases are introduced to the mean redshift, which is the $n(z)$'s leading-order statistic  affecting the cosmological constraints of cosmic shear analysis. We have verified this point using the \buzzard\ simulations; here we can marginalize over the effects of sample variance by producing multiple versions of the DES deep fields in different lines of sight. For a more detailed presentation of these results, we refer the reader to \citet{y3-sompz}. 

\subsection{Clustering redshifts}

We use clustering redshift methods to further constrain the $n(z)$ samples produced by 3sDIR. Clustering redshift methods exploit the 3D overlap between a target sample (the weak lensing sample) and a reference sample with accurate redshift estimates to infer the $n(z)$ of the former. Clustering redshift methods have been used in the past both to provide independent estimates of the $n(z)$ or to calibrate the mean of the redshift distributions obtained from photometric estimates \cite{Hildebrandt2017,Johnson2017,Davis2017aaa,Davis2018,Cawthon2018,Bates2019,vandenBusch2020,Hildebrandt2020}.

We fully describe the DES Y3 clustering methodology for the source sample in \cite{y3-sourcewz}.  We make use of two reference samples: red, luminous redMaGiC galaxies with high quality photo-$z$ estimates \citep{Rozo2016,y3-galaxyclustering} and spectroscopic galaxies from the BOSS and eBOSS surveys \citep{Smee2013,Dawson2013,Dawson2016,Ahumada2019}. The two samples complement each other: the former has a higher number density and covers the full DES Y3 footprint but it has a limited redshift coverage and it comes with a small uncertainty related to the redMaGiC photo-$z$  estimates. The latter is a spectroscopic sample and spans a larger redshift interval, allowing us to calibrate redshift distributions at higher redshift, but has a lower number density and  only $\sim 700$ sq. degrees of overlap with the DES Y3 footprint. Both samples have been reproduced in the simulations, and the BOSS and eBOSS selections are described in \citet{y3-sourcewz}.

For DES Y3, we explored two different approaches to using clustering redshift information. The first approach computes the mean redshift each tomographic bin in order to compare these estimates with the mean redshifts of the 3sDIR $n(z)$ samples. The mean redshift is only estimated in a limited redshift interval covered by the reference samples, excluding the tail of the distributions to mitigate the effect of magnification. This first approach is only used to cross-check the $n(z)$ from the 3sDIR. 

The second method includes the clustering redshift information in a likelihood analysis, joint with sample variance and shot noise from the 3sDIR method, that returns samples of probable redshift distributions, while marginalizing over a flexible model of the redshift evolution of source galaxy clustering bias, the dominant systematic in such clustering redshift studies. The second method generates an ensemble of redshift distributions used in the DES Y3 cosmological analysis \citep{y3-sompz}, and it is shown to vastly improve the accuracy of the shape of $n(z)$ derived from photometric data alone. Both methods have been tested in simulations and proved unbiased within uncertainties prior to application to data \citep{y3-sourcewz}. 

\subsection{Shear ratio}
\label{sec:shearratio}
In the DES Y3 cosmological analysis, we use the ratios of small-scale galaxy--galaxy lensing measurements sharing the same lens redshift bin and two different source tomographic bins to constrain redshift uncertainties and other systematics or nuisance parameters of our model. We briefly summarize this probe, and direct the reader to detailed description in \citet{y3-shearratio} for more information and robustness tests.

\emph{Lensing ratios} or \emph{shear ratios} have the advantage that, if the lens galaxies are tightly binned in redshift, they are mostly geometrical and can be modeled in the small, non-linear scale regime where we are not able to accurately model the original tangential shear quantity. Lensing ratios at small scales are therefore able to provide very valuable independent information that otherwise would be discarded. These shear-ratio measurements have been used before, especially as a test of the source redshift distributions, but this is the first time they are fully integrated as an additional probe within the \ttt\ project.
This permits us to use the entire constraining power of the lensing ratios, not only exploiting the dependency on the redshift distributions.

The shear-ratio data vector consists of nine numbers, each one corresponding to the scale-averaged lensing ratio for a given lens and two source bins. In this work, we use three lens redshift bins and four source redshift bins. Note that the DES Y3 \ttt\ project uses five lens bins, but we choose to discard the two highest redshift bins to construct lensing ratios both because they do not increase the S/N substantially and because the impact of lens magnification is much stronger for the highest redshift lens bins, and we prefer not to be dominated by lens magnification even though it is included in the modeling. From these redshift bins, given that we construct ratios of tangential shear measurements with a given fixed lens bin and two different source bins for each lens bin we can construct three independent ratios to make a total set of 9 independent ratios. Regarding the angular scales, we measure the lensing ratios in the same angular binning used for the \ttt\ analysis and then we apply the scale cuts as detailed in Table 2 of \citet{y3-shearratio}. Summarizing, we discard the scales already used in the \ttt\ analysis for the galaxy--galaxy lensing probe and scales where the IA model is not applicable for the ratio combinations that have significant overlap between source and lens galaxy redshift distributions, as these are most affected by IA. 

This data vector is used in a separate lensing ratio likelihood, assumed to be Gaussian. The covariance for this data vector comes from the propagation of the theoretical galaxy--galaxy lensing covariance and is independent of the \ttt\ covariance, as detailed and validated in \citet{y3-shearratio}. The constraints on the mean redshifts of each source tomographic bin from this likelihood are used as an independent validation of the SOMPZ and WZ estimates of these values. Furthermore, we combine this shear-ratio likelihood with the \ttt\ likelihood described below to self-consistently constrain source redshift distributions, as well as other parameters, such as shear multiplicative biases and intrinsic alignments. The \buzzard\ simulations were used to validate a number of assumptions made when using shear-ratio data, as described in \citet{y3-shearratio}, and we further validate the use of this information in this work showing that its combination with all other redshift information leads to unbiased cosmological constraints in our \buzzard\ simulated analyses. 

\section{Covariance Matrix}
\label{sec:covariance}

We model the statistical uncertainties of the two-point function measurements considered in this paper as a multivariate Gaussian distribution. Our model of the disconnected four-point function part of the covariance matrix of that distribution (also known as the Gaussian part of the covariance) is described in \citet{y3-covariances} and includes analytic treatment of bin averaging and sky curvature. The connected four-point function part of the covariance matrix and the contribution from super-sample covariance are modelled as described in \citet{Krause&Eifler2016}.

In \citet{y3-covariances} we demonstrate the robustness of our analysis with respect to the details of our covariance model and show that deviations from the Gaussian likelihood assumption are negligible in our analysis setup. In that paper, we identify approximations in our treatment of survey geometry to be the main source of inaccuracy in the covariance model. When considering the full \ttt\ data vector, this leads to an underestimation of our uncertainties in key cosmological parameters by about $3-4\%$ and on average it also increases the $\chi^2$ between measurements and the corresponding maximum likelihood models by about $4\%$.

We analytically marginalize over all enclosed mass parameters, which is possible under our assumptions of a Gaussian likelihood and a Gaussian prior for these parameters, characterized by a width of $\sigma_{B^{i}}$. We use very broad priors on these parameters, as motivated by \citet{y3-2x2ptbiasmodelling}. This would make our covariance numerically unstable to invert. In order to circumvent this, we use the Shermann--Morrison formula to analytically marginalize over $B^{i}$ directly in the inverse covariance matrix, $N^{-1}$, via:

\beq
N^{-1} = C^{-1}U(I + U^{T}C^{-1}U)^{-1}U^{T}C^{-1}\, ,
\eeq
where $C^{-1}$ is the inverse covariance matrix without the enclosed mass contribution, $I$ is the identity matrix, and $U$ is an $N_d\times N_{\rm lens}$ with $i$th column given by $\sigma_{B^{i}}\vec{t}^{i}$ and
\beq
t^{i}_a = \begin{cases} 
0 & \textrm{if lens redshift bin for element a is not $i$} \\
\beta_{ij}\theta_a & \textrm{otherwise}
\end{cases}
\eeq
where $a$ is the data vector index, $j$ is the source galaxy bin, and $\theta_a$ is the angular separation associated with the $a$th element of the data vector. We take $\sigma_{B^{i}}=10000$ for all $i$.

We generate the covariance used for all of the analyses in this work using true \buzzard\ cosmology, source and lens number densities, shape noise values, and redshift distributions.

\section{Redshift-space distortions in \buzzard}
\label{sec:rsd_bug}

In the process of preparing this paper for publication, an error in our implementation of redshift space distortions was discovered. Specifically, the proper motions of galaxies were included in their observed redshifts as 
\beq
z_{\rm obs} = z_{\rm cos} + \frac{v_{\rm LOS}}{c}
\eeq
instead of using the correct expression

\beq
z_{\rm obs} = z_{\rm cos} + (1 + z)\frac{v_{\rm LOS}}{c}
\eeq

where $z_{\rm cos}$ is the pure cosmological redshift, and $v_{\rm LOS}$ is the line-of-sight velocity. This error impacts the effect of redshift-space distortions on \wtheta\ and \gammat\ measurements. As our analysis was nearly completed when we discovered this error, we opted to correct for this effect in our model for RSD rather than change our simulations. This is accomplished by rescaling the linear growth rate that is used in the Kaiser model for RSD by a factor of $(1 + z)$, i.e. $f \rightarrow \frac{f}{(1+z)}$. 

The impact of this error on the analyses presented in this work is negligible: not accounting for this error increases the chi-squared for the analyses presented in \cref{sec:config_a} by 2.1 for \wtheta\ and the impact on all posteriors presented in this work is virtually unnoticeable.

\bibliographystyle{yahapj_twoauthor}
\bibliography{des_y1kp,des_y1kp_short,y3kp,adstex,wz}

\end{document}